\documentclass[aps,prl,twocolumn,letterpaper,nobibnotes,superscriptaddress,longbibliography]{revtex4-2}

\usepackage{amssymb}
\usepackage{amsmath}
\usepackage{graphicx}
\usepackage{extarrows}
\usepackage{url}
\usepackage{varioref}
\usepackage{hyperref}
\usepackage{cleveref}
\usepackage{lineno}
\usepackage{color}
\usepackage{ulem}
\usepackage{float}
\usepackage{verbatim}

\begin{document}

\title{Divergence of thermalization rates driven by the competition \\between finite
temperature and quantum coherence} 
\author{Yuqing Wang}
\affiliation{Department of Physics and State Key Laboratory of Low Dimensional Quantum Physics, Tsinghua University, Beijing, 100084, China}
\author{Libo Liang}
\affiliation{School of Electronics, Peking University, Beijing 100871, China}
\author{Qinpei Zheng}
\affiliation{School of Electronics, Peking University, Beijing 100871, China}
\author{Qi Huang}
\affiliation{School of Electronics, Peking University, Beijing 100871, China}
\author{Wenlan Chen}
\affiliation{Department of Physics and State Key Laboratory of Low Dimensional Quantum Physics, Tsinghua University, Beijing, 100084, China}
\affiliation{Frontier Science Center for Quantum Information and Collaborative Innovation Center of Quantum Matter, Beijing, 100084, China}
\author{Jing Zhang}
\affiliation{State Key Laboratory of Quantum Optics and Quantum Optics Devices, Institute of Opto-Electronics, Collaborative Innovation Center of Extreme Optics, Shanxi University, Taiyuan, Shanxi 030006, China}
\author{Xuzong Chen}
\affiliation{School of Electronics, Peking University, Beijing 100871, China}
\author{Jiazhong Hu}
\affiliation{Department of Physics and State Key Laboratory of Low Dimensional Quantum Physics, Tsinghua University, Beijing, 100084, China}
\affiliation{Frontier Science Center for Quantum Information and Collaborative Innovation Center of Quantum Matter, Beijing, 100084, China}
\affiliation{Beijing Academy of Quantum Information Science, Beijing, 100193, China}

\begin{abstract}
The thermalization of an isolated quantum system is described by quantum mechanics and thermodynamics, while these two subjects are still not fully consistent with each other.
This leaves a less-explored region where both quantum and thermal effects cannot be neglected, and the ultracold-atom platform provides a suitable and versatile testbed to experimentally investigate these complex phenomena. 
Here we perform experiments based on ultracold atoms in optical lattices and observe a divergence of thermalization rates of quantum matters when the temperature approaches zero. By ramping an external parameter in the Hamiltonian, we observe the time delay between the internal relaxation and the external ramping. 
This provides us with a direct comparison of the thermalization rates of different quantum phases. We find that the quantum coherence and bosonic stimulation of superfluid induces the divergence while the finite temperature and the many-body interactions are suppressing the divergence. 
The quantum coherence and the thermal effects are competing with each other in this isolated thermal quantum system, which leads to the transition of thermalization rate from divergence to convergence.
\end{abstract}
\maketitle

\section{I. introduction}
With the ramping of an external parameter, a classical isolated system evolves and attempts to thermalize into an equilibrium state at each specific number of the external parameter \cite{HuangBook}. 
If the ramping is very slow, the system will closely follow the ramping and remain in a steady state. 
However, as the ramping gets faster, the internal thermalization cannot keep pace with the change of the external parameter, resulting in a time delay of the observable values reaching to the corresponded one at the steady state under each external parameter. 

The scenario of a quantum isolated system at zero temperature is different.
The phenomenon is fully described by the coherent evolution of wave functions under quantum mechanics \cite{Ueda2020,Nandkishore2015,Abanin2019}, which is a different framework compared to thermodynamics. 
Nevertheless, as the particle number or system size increases, most systems can still exhibit thermalization described by the concept of entanglement thermalization \cite{Rigol2008,Trotzky2012,Kaufman2016,Wang2022}. 
In this framework, statistical physics can provide consistent predictions as quantum mechanics at zero temperature \cite{PhysRevLett.107.135301,PhysRevLett.121.170402,PhysRevLett.112.130403,Neill2016,PhysRevX.8.021030,SHENG202375}.

However, there remains a disparity between these two systems. 
In the case of a quantum system with finite temperature, the thermal effect interacts with and competes against quantum effects, but there have been few investigations in this context \cite{RevModPhys.83.863, ji2014experimental, eigen2018universal}. 
The quantum properties prevent the direct application of classical thermodynamics, and the finite temperature poses challenges in accurately calculating the quantum properties of these systems. 
This motivates us to use the ultracold-atom platform to investigate the thermalization behaviors under finite temperature, which provides a direct competition between quantum coherence and thermodynamics.
In our study, we note thermalization rate divergence near absolute zero temperature caused by boson stimulation, in contrast to the convergence triggered by finite temperature and many-body interactions.

\section{II. experimental setup}
Our system is based on the Bose-Hubbard model with ultracold atoms \cite{Bloch2012,science.aal3837}. 
In Fig.~1\textbf{a}, we present the phase diagram of the homogeneous Bose-Hubbard model as a function of the temperature $T$ and the ratio $g=J/U$ between tunneling amplitudes $J$ and the on-site interaction strength $U$. The value of $g$ can be adjusted by manipulating the trap depth of optical lattices \cite{Greiner2002}. 
At zero temperature, the system exhibits a quantum phase transition separating the superfluid and Mott insulator.
When the temperature is non-zero, the boundary of the superfluid shrinks which separates the superfluid and normal fluid with a continuous crossover \cite{PhysRevA.73.013408}.
The phase structure of superfluid/normal fluid bears similarities to the superconductor-metal transition.

\begin{figure}[t]
\includegraphics[width=\linewidth]{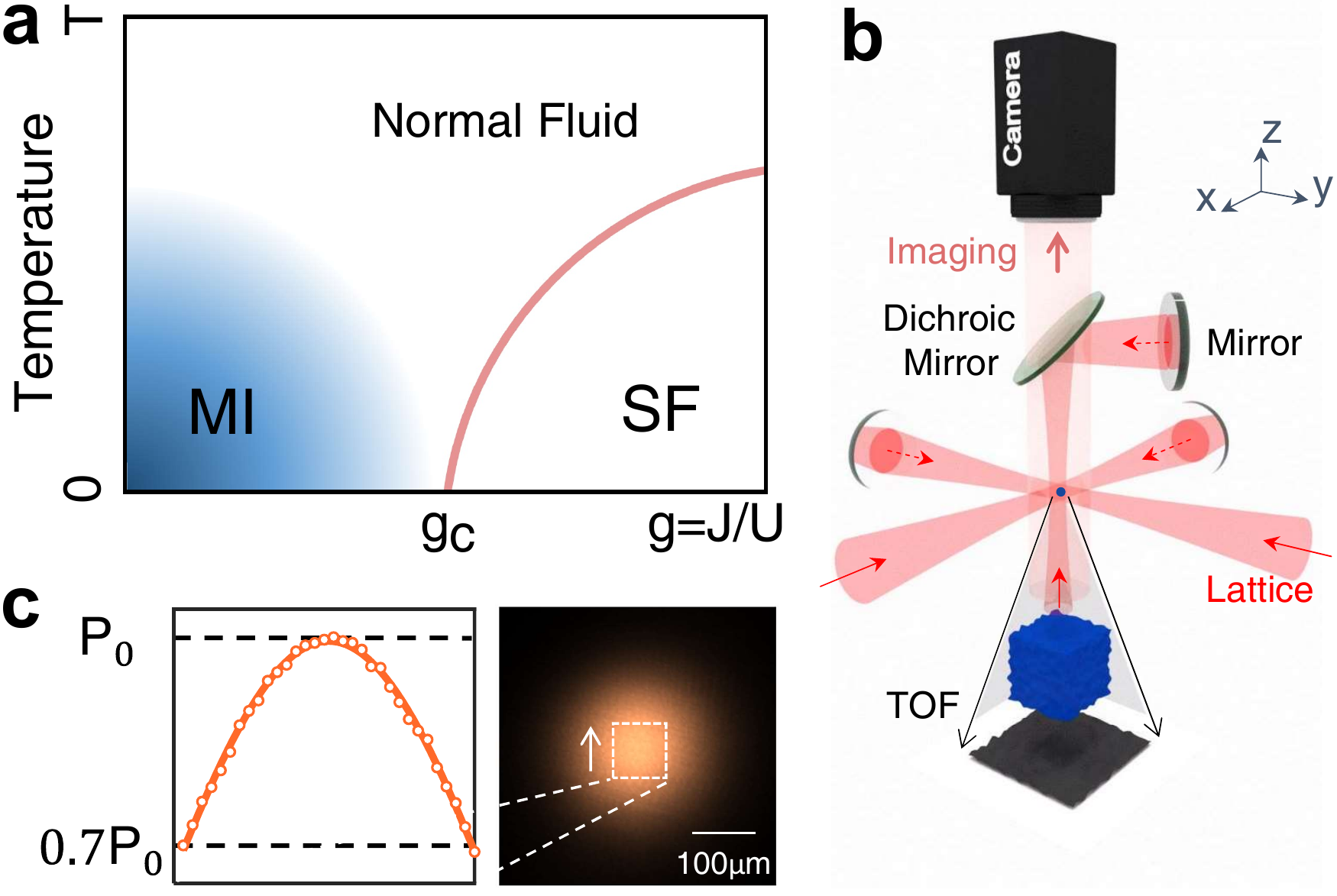}
\caption{\textbf{Phase diagram and the experimental setup.} \textbf{a} shows the phase diagram of cold atoms in optical lattices described by the homogeneous Bose-Hubbard model. The horizontal axis is labeled by the ratio $g=J/U$ between the tunneling $J$ and the on-site interaction strength $U$. The vertical axis is labeled by the temperature. When the temperature is zero, there is a quantum phase transition between the superfluid and Mott insulators at the critical point $g_c$.
We expect Mott-like quantum phenomena in the finite temperature region, and the name of Mott insulators is still used to label this region.
\textbf{b} shows the experimental setup of our optical-lattice system. Optical lattices are formed by three retro-reflected laser beams at a wavelength of 1064~nm. 
The band mapping is performed along the $z$ axis. \textbf{c} shows the measured intensity distribution of optical-lattice beams with a Gaussian shape and a waist 150~$\mu$m and there is no disorder in lattices}.
\end{figure}

Our measurement focuses on the ramping dynamics to investigate how the relaxation time $\tau_{SF}$ changes with the ramping speed $k$ at one particular parameter point $g_0$ under the influence of finite temperature effects. 
Inspired by the scaling relation in the Kibble-Zurek mechanism \cite{PhysRevB.72.161201,Dziarmaga2010,Campo2014}, we extend the concept of critical exponents to thermalization rate, characterizing the system's thermalization property and how it responds to external ramping.
\begin{equation}
\tau_{SF}\propto k^{-{\mathcal{C}\over 1+\mathcal{C}}}. \label{eq:time}
\end{equation}
In our experiment, we non-adiabatically change the parameter $g$ with a ramping speed $k$, and measure the relaxation time at selected point $g_0$. 
When the system exhibits fast thermalization, the adiabatic theorem applies, and atoms adiabatically follow the changes in $g$, resulting in the same relaxation time $\tau_{SF}$ as the time required to ramp the parameter, that is, $\tau_{SF} = |g_0-g_i|/k $ where $g_i$ is the initial value of the ramping parameter.
Consequently, $\mathcal{C}\rightarrow \infty$ approaches infinity for rapid thermalization. 
On the other hand, in the case of a system with a very slow thermalization rate, the observables do not track the external parameter ramping, and the relaxation time remains large but independent of ramping speed $k$, which leads to $\mathcal{C}\rightarrow 0$. 
This approach allows for a direct comparison of the intrinsic thermalization behaviors among different quantum phases with various $g_0$.

We utilize the improved band mapping technique  \cite{PhysRevLett.127.200601,PhysRevLett.94.080403,LIANG20222550,PhysRevResearch.5.013136} to measure the quasi-momentum distribution of the system. 
This allows us to determine the number of atoms that exhibit coherence within the first Brillouin zone. 
This distribution directly corresponds to the phase coherence associated with superfluidity, which serves as an order parameter distinguishing superfluid and incoherent phases.
We define the relaxation time $\tau_{SF}$ as the duration required to attain the same coherence level in an adiabatic case, and obtain the thermalization rate $\mathcal{C}$.
This process is repeated for atoms with the same atom number $N_0$ but different initial temperatures $T_0$.

\begin{figure}[t]
\includegraphics[width=\linewidth]{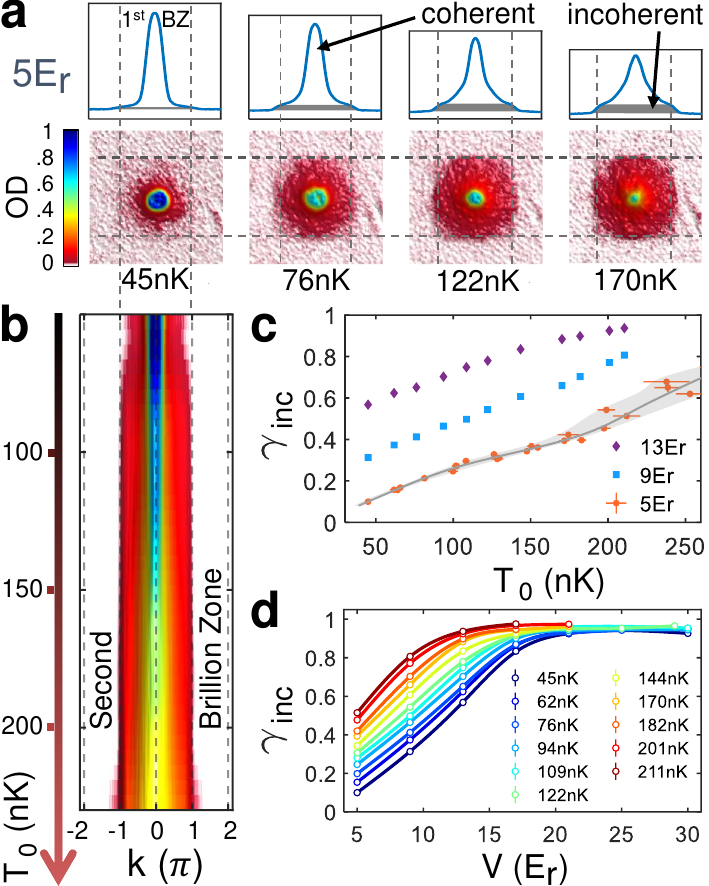}
\caption{\textbf{Band mapping at different temperatures $T_0$.} \textbf{a} shows a few examples of the quasi-momentum distribution both along $x$ direction and in $x$-$y$ plane with different initial BEC temperature $T_0$ at the lattice trap depth $5E_r$. The dashed lines label out the first Brillouin zone. The shadow area labels the incoherent parts in the quasi-momentum distribution and this helps us to exact out the incoherent fraction $\gamma_{inc}$.
\textbf{b} shows how the quasi-momentum distribution changes versus $T_0$ in the first and second Brillouin zones at the lattice trap depth $5E_r$. The vertical axis is the initial temperature $T_0$. We find that the second Brillouin zone will be excited when $T_0$ is larger than 220~nK.
\textbf{c} shows $\gamma_{inc}$ versus $T_0$ at different trap depth (5$E_r$, 9$E_r$, and 13$E_r$). 
The error bars correspond to one standard deviation.
\textbf{d},~we plot $\gamma_{inc}$ versus the lattice trap depth $V$ at $T_0$ of 45, 62, 76, 94, 109, 122, 144, 170, 182, 201, and 211~nK. 
}
\end{figure}

We prepare a total of $N_0=1.3(1)\times 10^5$ rubidium-87 atoms in condensates within a dipole trap, where the initial temperature $T_0$ can be adjusted.
Here $T_0$ is measured by characterizing the thermal component of condensates (See Supplementary materials Section 2.).
The dipole trap is a crossed dipole trap with the lowest vibrational frequencies of 42 and 60~Hz along horizontal and vertical directions.
Then we adiabatically transfer the atoms from the dipole trap to the three-dimensional optical lattices, where the optical lattices are formed by the interference of three retro-reflecting laser beams with a wavelength at $\lambda=1064$~nm. 
The dipole trap is fully turned off when the optical lattices are turned on, and the trap depth of lattices is set at 5$E_r$ where $E_r=h^2/2m\lambda^2=h\times 2$~kHz is the recoil energy.
This transfer process takes approximately 80~ms.
Once loaded into the lattices, the atoms have a radius of around 10~$\mu$m.
Subsequently, we hold the atoms in the lattices for 100~ms. 
The Gaussian beam waist of each lattice beam is 150~$\mu$m.
As a result, the system is described by a Bose-Hubbard model with a harmonic-trap background, as \cite{Sherson2010,PhysRevB.40.546}

\begin{eqnarray}
H_0&=&-J\sum_{\langle i,j\rangle}(a^\dagger_i a_j+a^\dagger_j a_i)+{1\over 2}U\sum_i n_i(n_i-1) \nonumber \\
& &+\sum_i({1\over2} m\omega^2 r^2_i-\mu)n_i.
\end{eqnarray}
Here $a_i$ represents the annihilation operator of one particle at lattice site $i$, and $n_i=a^\dagger_i a_i$ is the particle number operator. The Gaussian shape of the lattice beams provides an isotropic harmonic trap, and the vibrational frequency $\omega$ is $2\pi\times$44~Hz at the trap depth $V=5E_r$. The frequency $\omega$ scales as $\omega\propto \sqrt{V}$.
Due to the Gaussian envelope of the lattice beams, the variation of $J/U$ for atoms on the outer regions compared to the center part is only 1\% difference \cite{PhysRevA.72.053606,Liang23}.
    
\section{III. divergence of thermalization rate}
Firstly, we calibrate our system (Fig.~2\textbf{a}) \cite{PhysRevLett.127.200601,PhysRevLett.94.080403,LIANG20222550,PhysRevResearch.5.013136} to distinguish the different phase components and determine the number of incoherent atoms. 
We employ the relative incoherent fraction $\gamma_{inc}$ to characterize the system. 
Since the direct measurement of the real temperature in optical lattices is challenging \cite{carcy2021certifying}, we use the initial temperature $T_0$ as a characterization parameter instead.
In Fig.~2\textbf{b}, we present the data of band mapping as a function of $T_0$ at the trap depth $V=5E_r$. As $T_0$ increases, the thermal effects become more dominant, resulting in a higher fraction of incoherent atoms. However, when $T_0$ exceeds 220~nK, atoms are excited into the second Brillouin zone, where the incoherent fraction becomes difficult to extract, which is what we want to avoid. At around 50~nK, almost all the atoms remain coherent and stay in the superfluid phase at a trap depth of $5E_r$. 
In addition, we also calibrate the change of the incoherent fraction $\gamma_{inc}(T_0)$ at different points $g_0$ corresponding to different trap depth $V$ (Fig.~2\textbf{c} and \textbf{d}).
Here when $V$ is above $13E_r$, the Mott insulators start to gradually appear for Rubidium-87 atoms \cite{Greiner2002,Liang23} in such inhomogeneous optical lattices. Based on these calibrations, we obtain the steady incoherent fraction $\gamma_{inc}(T_0)$ versus the initial temperature $T_0$ and different trap depth $V$.

\begin{figure*}[t]
\includegraphics[width=\linewidth]{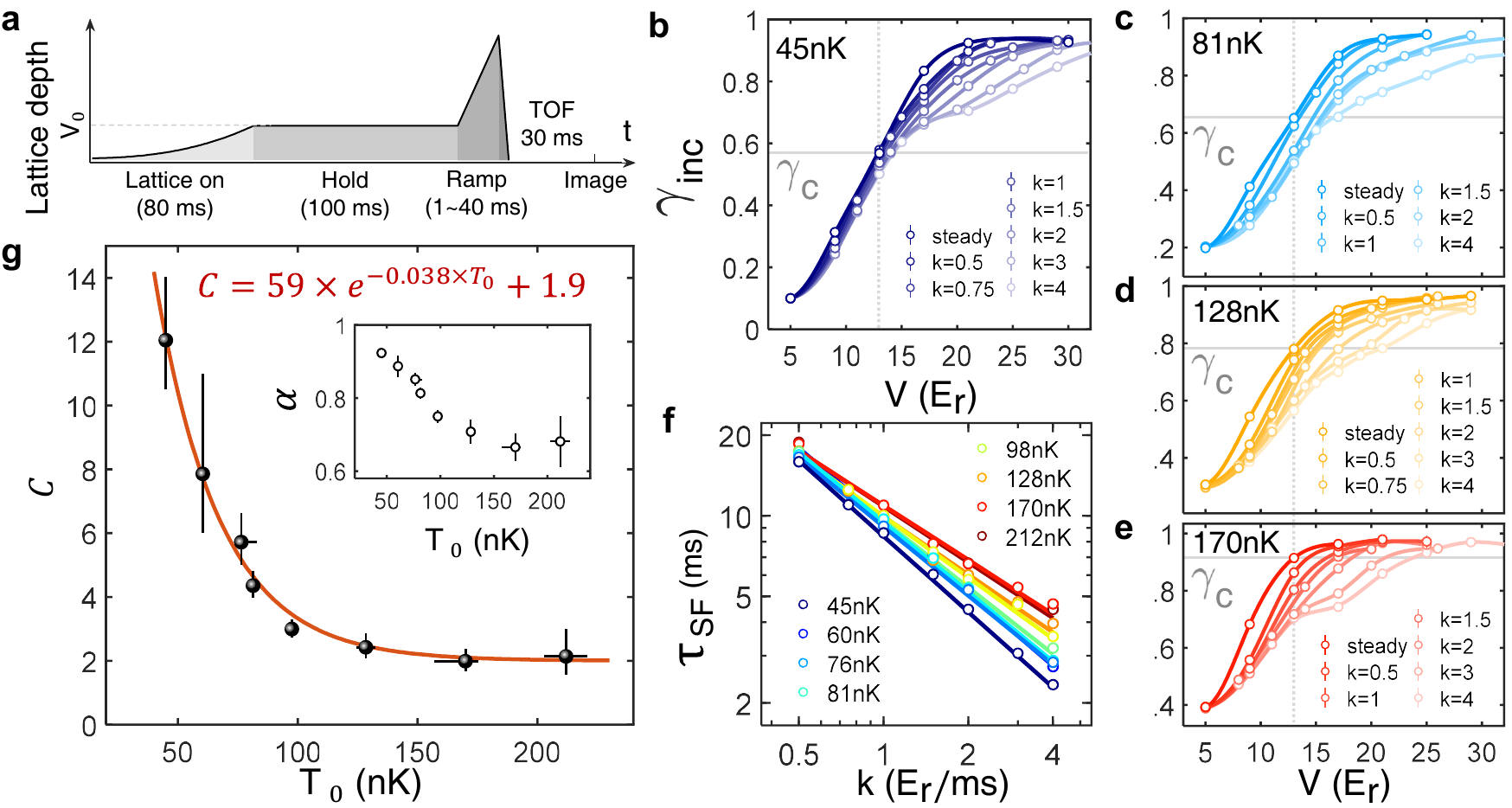}
\caption{\textbf{The relaxation measurement and the divergence of thermalization rates.} \textbf{a} shows the time sequence. We prepare superfluid at 5$E_r$ and then linearly ramp up the trap depth. Once reaching the desired trap depth $V$, we ramp down the $x$ and $y$ lattices in 2~ms and turn off the $z$ lattice immediately. After that, a 30~ms time-of-flight is performed to measure the quasi-momentum distribution.
\textbf{b} to \textbf{e} show the incoherent fraction $\gamma_{inc}$ versus the ramping speed $k$ and the trap depth $V$ at the initial temperature $T_0=$45, 81, 128, and 170~nK. 
The unfilled circles are the data points and the solid lines are polynomial fit.
In each panel with a fixed $T_0$, we extract how long it takes for the system to reach the steady incoherent fraction $\gamma_c(T_0)$ at the selected point 13$E_r$. The vertical dashed gray line labels out 13$E_r$, and the solid horizontal gray line labels out the critical fraction $\gamma_c(T_0)$ for different $T_0$ in each panel. This helps us to extract the relaxation time of superfluid at different ramping speeds and temperatures.
\textbf{f} shows the relaxation time versus ramping speeds and temperatures. The unfilled circles are the data points, and each temperature $T_0$ is labeled by different colors. The solid lines are scaling fits of $\tau_{SF}\propto k^\alpha$ for each particular $T_0$. Based on the scaling relations, we obtain the thermalization rate $\mathcal{C}$. \textbf{g} shows $\mathcal{C}$ versus $T_0$. The bottom horizontal axes are labeled by the initial condensate temperature $T_0$. When the temperature approaches zero, $\mathcal{C}$ increases significantly. The inset shows the fitting exponent $\alpha$ versus $T_0$. All the error bars correspond to one standard deviation.
}
\end{figure*}

Next, we perform the ramping of the trap depth $V$ linearly with a ramping speed $k$ for samples with different initial temperatures $T_0$ (Fig.~3\textbf{a}). At each specific $V$, we perform the band mapping and extract $\gamma_{inc}(V,T_0,k)$ (Fig.~3\textbf{b} to \textbf{e}). During this process, we ensure that there are no excitations into the second Brillouin zone when controlling $T_0$ below 220~nK. 
When the temperature is low (45~nK in Fig.~3\textbf{b}), there is a clear turning point at 13$E_r$ for different ramping speed $k$. The dynamical responses are almost the same before the critical point $13E_r$ and start to furcate after it.
Therefore, we utilize the steady incoherent fraction $\gamma_{inc}(T_0)$ at the critical point $13E_r$ as a reference to measure the time required to reach the same level of incoherence.
This procedure provides us with a relation time $\tau_{SF}(T_0,k)$ for each ramping speed $k$ and $T_0$. By plotting the logarithmic relation of $\tau_{SF}(T_0,k)$ versus $k$ in Fig.~3\textbf{f} and fitting the data using the scaling relation $\tau_{SF}(T_0,k)\propto k^{-\alpha}$, we determine the value of $\alpha$, which is related to the thermalization rate $\mathcal{C}$ through $\mathcal{C}=\alpha/(1-\alpha)$.
particularly, the critical exponent $\nu z$ can be obtained via $\nu z=\mathcal{C}$ at the critical point.
In the logarithmic plot, the fitting lines exhibit good agreement with the data, confirming the validity of the scaling relation.
Specifically, at $T_0=45$~nK, $\alpha$ is 0.92(1) corresponding to $\mathcal{C}=12.05^{_{+1.99}}_{^{-1.52}}$.
At $T_0=76$~nK, $\alpha$ is 0.85(2) and the corresponded $\mathcal{C}$ is $5.72^{_{+0.92}}_{^{-0.72}}$.
At $T_0=128$~nK, $\alpha$ is 0.71(3) and the corresponded $\mathcal{C}$ is $2.42^{_{+0.45} }_{^{-0.35}}$.
In Fig.~3\textbf{g}, we plot the relation between $\mathcal{C}$ and $T_0$, while the inset panel shows the plot of $\alpha$ versus $T_0$. It is evident that $\mathcal{C}$ increases dramatically as the temperature approaches zero. We fit this change using a trial function, which only serves as an empirical fit based on the data
\begin{equation}
\mathcal{C}=59(9)\times \exp(-0.038(3)\times T_0)+1.9(3).
\end{equation}

We think the superfluid state itself should bear the responsibility of divergence instead of any parameters in Hamiltonian since only incoherence can converge the thermalization rates.
In superfluid, most of the atoms condense into the zero-momentum state, and this leads to the bosonic stimulation \cite{Miesner1998,Lu2023} that the excitation speed from superfluid will be amplified by the condensed atoms with a factor $\sqrt{\tilde{N}}$. Here $\tilde{N}$ is the condensed-atom number. 
When the temperature increases, the increasing proportion of incoherence attenuates the bosonic stimulations and quantum coherence, and then the thermalization rates converge.

One way to further support this argument is to inspect the thermalization rates of Mott insulators. Even at zero temperature, the superfluid will diminish when the interaction dominates the system, and this will also suppress the bosonic stimulation.
Therefore, we examine the thermalization rate $\mathcal{C}$ using different truncation trap depth.
In Fig.~4, we illustrate the variation of $\mathcal{C}$ as a function of trap depth for different temperatures. 
We observe that the divergence of $\mathcal{C}$ exhibits a similar pattern to the phase diagram.
Specifically, the superfluid phase is associated with larger $\mathcal{C}$ in the regime with lower $T_0$ and smaller $V$. In the Mott insulator or normal fluid regime, $\mathcal{C}$ rapidly decays to around 2.
This suggests that the divergence of $\mathcal{C}$ is an intrinsic behavior of superfluid due to the quantum coherence instead of disorders.

\begin{figure}[bt]
\includegraphics[width=\linewidth]{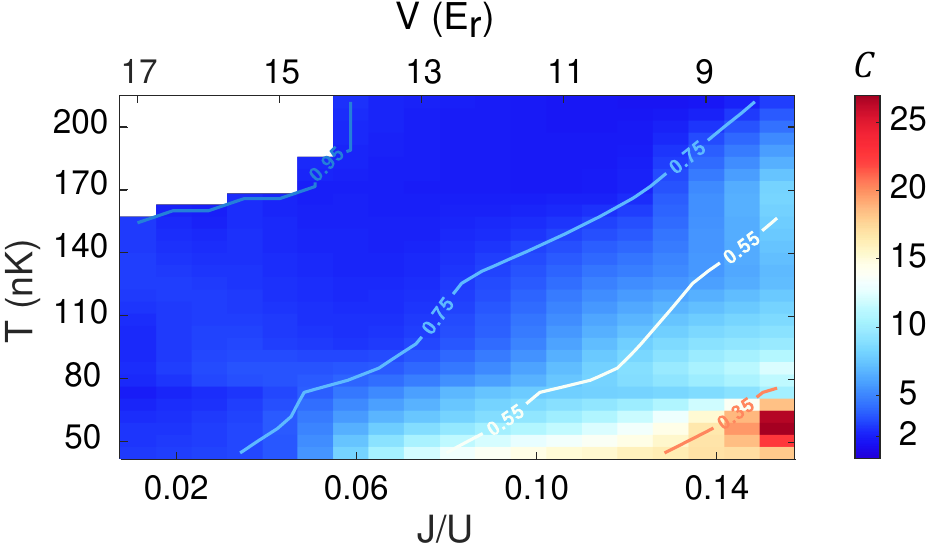}
\caption{\textbf{Thermalization $\mathcal{C}$ with different $T$ and trap depth $V$.} The colored background represents the distribution of thermalization rates, with contour lines indicating incoherent fractions as denoted by various $\gamma_{inc}$ values (0.35, 0.55, 0.75, and 0.95), thereby providing a clear view of superfluid density. In regions of shallow traps and near-zero temperatures, $\mathcal{C}$ tends towards divergence (bright yellow or red region). Conversely, as the trap deepens towards a critical depth at zero temperature and the temperature rises, $\mathcal{C}$ converges to around 2  (deep blue region) with $\gamma_{inc}$ approaching a value near 1. Here we do not calculate $\mathcal{C}$ for $\gamma_{inc}$ larger than 0.95 due to the measurement accuracy.
}
\end{figure}

\section{IV. conclusion}
In conclusion, our study examines the thermalization behaviors of cold bosonic systems in optical lattices at various temperatures. 
Based on the scaling relations of relaxation time versus the external parameter ramping rate, we extract the thermalization rate which exhibits a divergence as the temperature approaches zero, reflecting the intrinsic stimulation of the superfluid phase. 
Conversely, the thermalization rate converges as either the temperature or interaction increases, where the phase decoherence or thermal excitation suppresses the speeding up. This is a direct competition between quantum and thermal effects. 
We believe our work provides fresh perspectives on quantum systems at finite temperature and related to the divergent \cite{10_Ying2015Science,PhysRevLett.127.137001,Reiss2021,Wang_2024} or scaling \cite{PhysRevLett.95.105701,Navon167,CChinKBMZ,Ko2019,LukinKBMZ,PhysRevLett.125.260603} behaviors in other quantum materials.

\section{acknowledgments}
We acknowledge the stimulating discussions from Zhiyuan Yao.
This work is supported by National Key Research and Development Program of China (2021YFA0718303, 2021YFA1400904), National Natural Science Foundation of China (92165203, 61975092, 11974202), and Tsinghua University Initiative Scientific Research Program.

\section{Appendix A: extraction of $\gamma_{inc}$ from the band mapping}
In the experiment, we use the improved band-mapping method to measure the incoherent fraction $\gamma_{inc}$. The band-mapping method is an efficient approach to measuring the quasi-momentum distribution for different phases. By adiabatically ramping down the power of all lattice beams, the lattice quasi-momentum is converted to real-space momentum. Nonetheless, during this process, the momentum distribution is disturbed by the on-site interaction, thus we cannot obtain the ideal uniform distribution for Mott insulators. In our improved method, unlike the slowly ramped $x$ and $y$ lattices, the $z$ lattice is turned off immediately to release the interaction, and this gives us a better converted real-space momentum.

In the deep Mott region, the wavefunction of each atom is a Wannier function localized within one site and is also a superposition of all possible Bloch functions within one Brillouin zone. Therefore, the quasi-momentum is uniformly distributed as a flat plateau in the first Brillouin zone. On the contrary, the superfluid phase corresponds to a central peak since the atoms tend to be in the zero-momentum state. We define the incoherent fraction as:
\begin{equation}
    \gamma_{inc}=1-\frac{A_{pk}}{A_{tot}}
\end{equation}
where $A_{pk}$ denotes the area of the central coherent peak, and $A_{tot}$ is the total area of the first Brillouin zone. In our experiment, the phase decoherence leads to the increase of incoherent atoms, thus the increase of $\gamma_{inc}$.

In order to eliminate the impact of in-alignment errors, the quasi-momentum distribution is firstly centralized and symmetrized. We choose an approximate central momentum and turn the right momentum distribution to the left, then calculate the residual error in a chosen area with the unprocessed data. The central point is swept in a reasonable range to find the optimal point where the residuals are minimal. After the centralization, the coherent peak is easy to be distinguished based on the plateau.

\begin{figure}[htbp]
\includegraphics[width=0.9\linewidth]{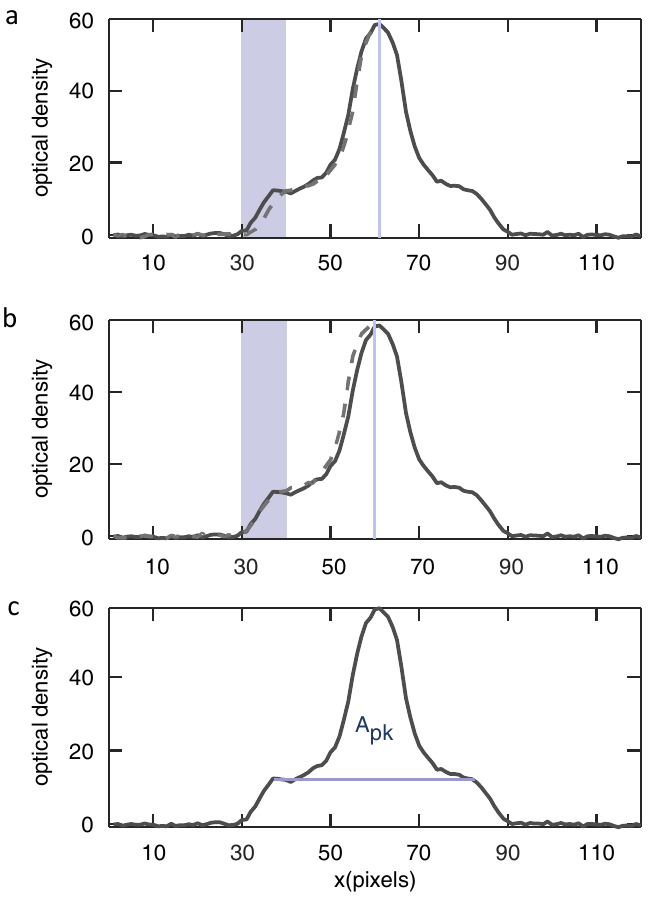}
\centering
\caption{\textbf{Extraction of $\gamma_{inc}$ from quasi-momentum distribution.} Panel \textbf{a} and \textbf{b} show the centralization and symmetrization before calculating $\gamma_{inc}$. The quasi-momentum distribution is shown as solid line, while the dashed line is the part that flips from right to left along the center (purple line). We calculate the residuals in the purple area between two lines, which is the criterion for choosing the central point. In panel \textbf{c}, the coherent part is labeled as $A_{pk}$, and $\gamma_{inc}$ is the fraction of the flat plateau in the total area. }
\end{figure}

\section{Appendix B: Measuring the condensates' temperature}
In this section, we show how we measure the initial condensate temperature $T_0$.
To measure $T_0$, we release the condensate from the optical dipole trap for a 30-ms-long free expansion, and then apply a resonant absorption imaging.
Based on the time-of-flight method, we obtain the two-dimensional density $n(x,y)$. Then, we integrate one direction and make the one-dimensional density $n(x)$ to be fitted by a bimodal distribution \cite{PhysRevLett.99.120404, PhysRevB.75.134302},
\begin{eqnarray}
    n(x)&=&n_{th}(x)+n_{BEC}(x) \nonumber\\
    &=&n_{th}^0\times \frac{g_{5/2}\left(\exp\left[-(\frac{x-x_0}{\sigma_{th}})^2\right]\right)}{g_{5/2}(1)}+ \nonumber\\
    &&n_{BEC}^0\times \left(\max\left\{1-(\frac{x-x_0}{\sigma_{BEC}})^2,0\right\}\right)^{2},
\end{eqnarray}
where $x_0$ stands for the central place of atoms, $n_{th}^0$ and $n_{BEC}^0$ are the central densities of thermal atoms and condensates respectively, $\sigma_{th}$, $\sigma_{BEC}$ are the radii of thermal atoms and condensates, and $g_\alpha(x)=\sum_nz^n/n^\alpha$ is the Bose function.

Then we calculate the condensate fraction $N_0/N$ using the fitting results above and the critical phase transition temperature $T_c$ by the formula 
\begin{equation}
    k_BT_c=0.94\hbar\omega N^{1/3},
\end{equation}
where $\omega$ is the vibrational frequency of the trap, $N$ is the total atom number, and $k_B$ is the Boltzmann constant. 
According to the relation of the condensate fraction and the temperature
\begin{equation}
    \frac{N_0}{N}=1-(\frac{T}{T_c})^a,
\end{equation}
where $a$ equals to 3 for the harmonic trap, we obtain the temperature corresponding to each experimental measurement.

\section{Appendix C: Fitting procedure}

The fitting process consists of two steps. In the first step, the data of $\gamma_{inc}$ under the same temperature is interpolated into smooth curves. Then we take $\gamma_{inc}$ at the critical point as a criterion to extract the relaxation time. The details are described in the following subsections.

\subsection{A. fitting by interpolation}

For each set of $T_0$ and $k$, we plot the incoherent fraction $\gamma_{inc}(V,T_0,k)$ as a function of the trap depth $V$ and fit the data via cubic spline interpolation. 
The initial state is prepared at $V=5E_r$ for all different $k$, so the starting points of different $k$ under the same $T_0$ are the same point.
As shown in Fig.~6, the fitted curves corresponding to different ramp speed $k$ start at the same point and then diverge slowly as $V$ increases.

\begin{figure}[htbp]
\includegraphics[width=0.9\linewidth]{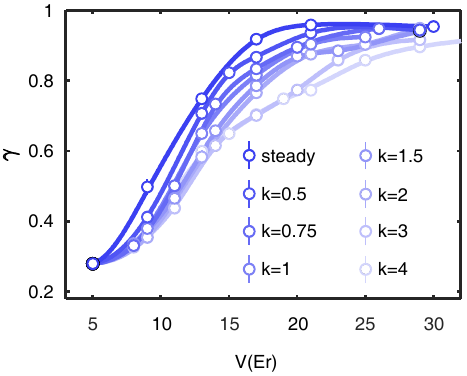}
\centering
\caption{\textbf{Interpolation curve of $\gamma_{inc}$ versus $k$.} All fitted curves start at the same point at $V=5E_r$.
}
\end{figure}

\subsection{B. Extraction of the relaxation time $\tau_{SF}$}

Rather than being capable of adiabatic evolution, the system is unable to respond instantaneously to the external ramp, necessitating additional relaxation time. 
For instance, we adopt the equilibrium incoherent fraction, denoted as $\gamma_{c}$, at the critical point of $V=13E_r$ as a reference, and we define the relaxation time $\tau_{SF}$ as the time required to achieve $\gamma_{c}$. 
Practically, for specific temperature, we delineate a vertical dashed line at $V=13E_r$ and subsequently draw a horizontal line at $\gamma_{c}$, which is the vertical coordinate at the intersection of the dashed line and the steady-state curve. 
We can then determine the trap depth from the intersections of this horizontal line with different ramping curves for $k$. 
Given the known ramping speed, we derive the relaxation time, and perform the thermalization fitness as shown in the main text.

We can also select other reference trap depths to elucidate the correlation between superfluid density and the thermalization rate. 
In this case, the critical point $V=13E_r$ is substituted by $V^{\prime}$, and the corresponding equilibrium incoherent fraction is denoted as $\gamma_{c}^{\prime}$. 
The subsequent procedures remain the same as for the critical one, but these yield the thermalization rate under varying trap depths and temperatures.

\begin{figure}[htbp]
\includegraphics[width=0.9\linewidth]{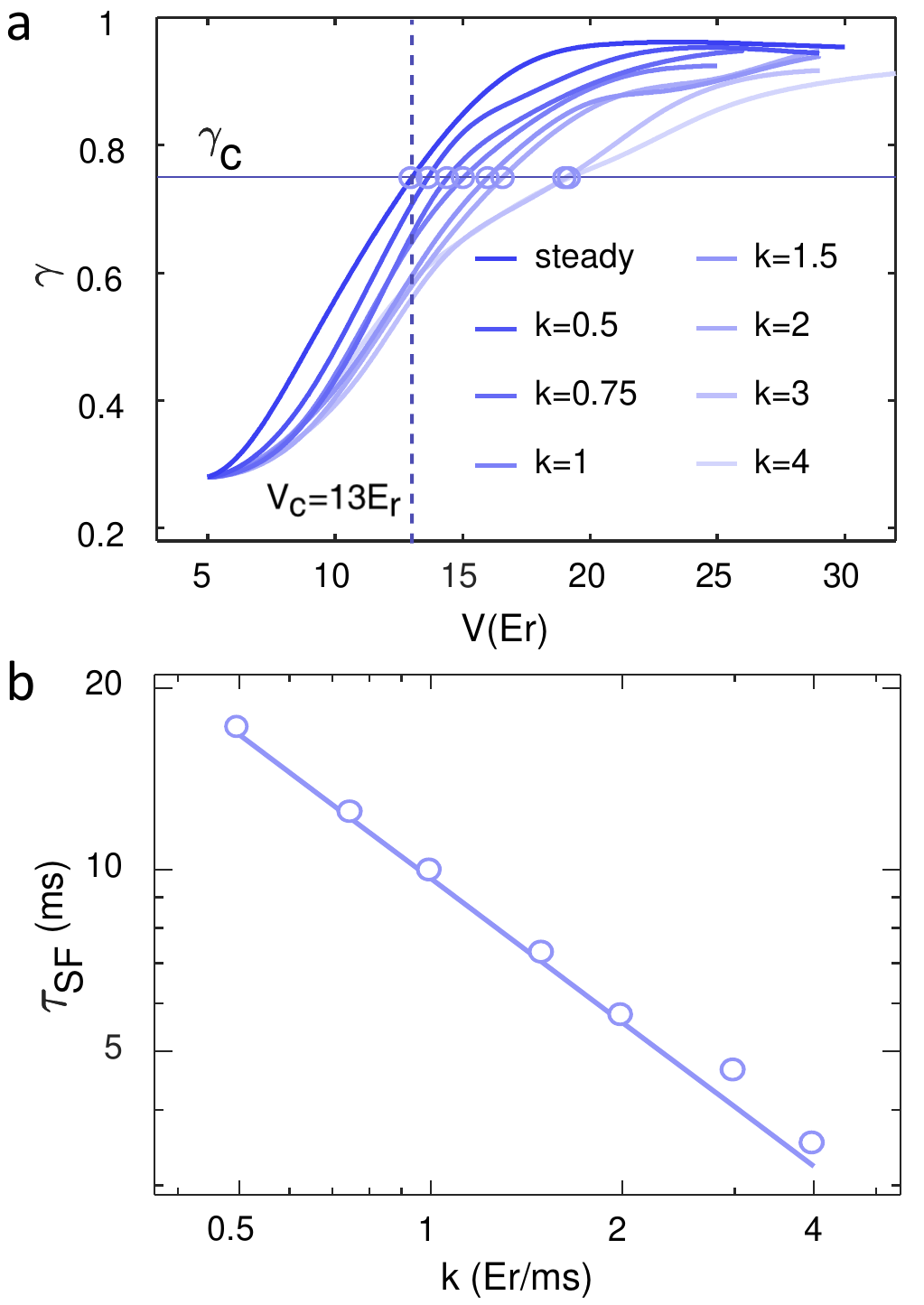}
\centering
\caption{\textbf{Relaxation time extracted from Fig.~6.} The vertical dashed line goes through the critical point $V_c=13E_r$, providing us the critical incoherent fraction $\gamma_c$ from the intersection with the steady curve. The solid line goes through $\gamma_c$, and the relaxation time is extracted from all the intersections with the fitted curves.}
\end{figure}


\begin{thebibliography}{47}%
\makeatletter
\providecommand \@ifxundefined [1]{%
    \@ifx{#1\undefined}
}%
\providecommand \@ifnum [1]{%
    \ifnum #1\expandafter \@firstoftwo
    \else \expandafter \@secondoftwo
    \fi
}%
\providecommand \@ifx [1]{%
    \ifx #1\expandafter \@firstoftwo
    \else \expandafter \@secondoftwo
    \fi
}%
\providecommand \natexlab [1]{#1}%
\providecommand \enquote  [1]{``#1''}%
\providecommand \bibnamefont  [1]{#1}%
\providecommand \bibfnamefont [1]{#1}%
\providecommand \citenamefont [1]{#1}%
\providecommand \href@noop [0]{\@secondoftwo}%
\providecommand \href [0]{\begingroup \@sanitize@url \@href}%
\providecommand \@href[1]{\@@startlink{#1}\@@href}%
\providecommand \@@href[1]{\endgroup#1\@@endlink}%
\providecommand \@sanitize@url [0]{\catcode `\\12\catcode `\$12\catcode `\&12\catcode `\#12\catcode `\^12\catcode `\_12\catcode `\%12\relax}%
\providecommand \@@startlink[1]{}%
\providecommand \@@endlink[0]{}%
\providecommand \url  [0]{\begingroup\@sanitize@url \@url }%
\providecommand \@url [1]{\endgroup\@href {#1}{\urlprefix }}%
\providecommand \urlprefix  [0]{URL }%
\providecommand \Eprint [0]{\href }%
\providecommand \doibase [0]{https://doi.org/}%
\providecommand \selectlanguage [0]{\@gobble}%
\providecommand \bibinfo  [0]{\@secondoftwo}%
\providecommand \bibfield  [0]{\@secondoftwo}%
\providecommand \translation [1]{[#1]}%
\providecommand \BibitemOpen [0]{}%
\providecommand \bibitemStop [0]{}%
\providecommand \bibitemNoStop [0]{.\EOS\space}%
\providecommand \EOS [0]{\spacefactor3000\relax}%
\providecommand \BibitemShut  [1]{\csname bibitem#1\endcsname}%
\let\auto@bib@innerbib\@empty
\bibitem [{\citenamefont {Huang}(2009)}]{HuangBook}%
    \BibitemOpen
    \bibfield  {author} {\bibinfo {author} {\bibfnamefont {K.}~\bibnamefont {Huang}},\ }\href {https://doi.org/10.1201/9781439878132} {\emph {\bibinfo {title} {{Introduction to Statistical Physics (2nd ed.)}}}}\ (\bibinfo  {publisher} {Chapman and Hall/CRC},\ \bibinfo {year} {2009})\BibitemShut {NoStop}%
\bibitem [{\citenamefont {Ueda}(2020)}]{Ueda2020}%
    \BibitemOpen
    \bibfield  {author} {\bibinfo {author} {\bibfnamefont {M.}~\bibnamefont {Ueda}},\ }\bibfield  {title} {\bibinfo {title} {{Quantum equilibration, thermalization and prethermalization in ultracold atoms}},\ }\href {https://doi.org/10.1038/s42254-020-0237-x} {\bibfield  {journal} {\bibinfo  {journal} {Nat. Rev. Phys.}\ }\textbf {\bibinfo {volume} {2}},\ \bibinfo {pages} {669} (\bibinfo {year} {2020})}\BibitemShut {NoStop}%
\bibitem [{\citenamefont {Nandkishore}\ and\ \citenamefont {Huse}(2015)}]{Nandkishore2015}%
    \BibitemOpen
    \bibfield  {author} {\bibinfo {author} {\bibfnamefont {R.}~\bibnamefont {Nandkishore}}\ and\ \bibinfo {author} {\bibfnamefont {D.~A.}\ \bibnamefont {Huse}},\ }\bibfield  {title} {\bibinfo {title} {Many-body localization and thermalization in quantum statistical mechanics},\ }\href {https://doi.org/10.1146/annurev-conmatphys-031214-014726} {\bibfield  {journal} {\bibinfo  {journal} {Annu. Rev. Condens. Matter Phys.}\ }\textbf {\bibinfo {volume} {6}},\ \bibinfo {pages} {15} (\bibinfo {year} {2015})}\BibitemShut {NoStop}%
\bibitem [{\citenamefont {Abanin}\ \emph {et~al.}(2019)\citenamefont {Abanin}, \citenamefont {Altman}, \citenamefont {Bloch},\ and\ \citenamefont {Serbyn}}]{Abanin2019}%
    \BibitemOpen
    \bibfield  {author} {\bibinfo {author} {\bibfnamefont {D.~A.}\ \bibnamefont {Abanin}}, \bibinfo {author} {\bibfnamefont {E.}~\bibnamefont {Altman}}, \bibinfo {author} {\bibfnamefont {I.}~\bibnamefont {Bloch}},\ and\ \bibinfo {author} {\bibfnamefont {M.}~\bibnamefont {Serbyn}},\ }\bibfield  {title} {\bibinfo {title} {Colloquium: Many-body localization, thermalization, and entanglement},\ }\href {https://doi.org/10.1103/RevModPhys.91.021001} {\bibfield  {journal} {\bibinfo  {journal} {Rev. Mod. Phys.}\ }\textbf {\bibinfo {volume} {91}},\ \bibinfo {pages} {021001} (\bibinfo {year} {2019})}\BibitemShut {NoStop}%
\bibitem [{\citenamefont {Rigol}\ \emph {et~al.}(2008)\citenamefont {Rigol}, \citenamefont {Dunjko},\ and\ \citenamefont {Olshanii}}]{Rigol2008}%
    \BibitemOpen
    \bibfield  {author} {\bibinfo {author} {\bibfnamefont {M.}~\bibnamefont {Rigol}}, \bibinfo {author} {\bibfnamefont {V.}~\bibnamefont {Dunjko}},\ and\ \bibinfo {author} {\bibfnamefont {M.}~\bibnamefont {Olshanii}},\ }\bibfield  {title} {\bibinfo {title} {{Thermalization and its mechanism for generic isolated quantum systems}},\ }\href {https://doi.org/10.1038/nature06838} {\bibfield  {journal} {\bibinfo  {journal} {Nature}\ }\textbf {\bibinfo {volume} {452}},\ \bibinfo {pages} {854} (\bibinfo {year} {2008})}\BibitemShut {NoStop}%
\bibitem [{\citenamefont {Trotzky}\ \emph {et~al.}(2012)\citenamefont {Trotzky}, \citenamefont {Chen}, \citenamefont {Flesch}, \citenamefont {McCulloch}, \citenamefont {Schollw{\"{o}}ck}, \citenamefont {Eisert},\ and\ \citenamefont {Bloch}}]{Trotzky2012}%
    \BibitemOpen
    \bibfield  {author} {\bibinfo {author} {\bibfnamefont {S.}~\bibnamefont {Trotzky}}, \bibinfo {author} {\bibfnamefont {Y.-A.}\ \bibnamefont {Chen}}, \bibinfo {author} {\bibfnamefont {A.}~\bibnamefont {Flesch}}, \bibinfo {author} {\bibfnamefont {I.~P.}\ \bibnamefont {McCulloch}}, \bibinfo {author} {\bibfnamefont {U.}~\bibnamefont {Schollw{\"{o}}ck}}, \bibinfo {author} {\bibfnamefont {J.}~\bibnamefont {Eisert}},\ and\ \bibinfo {author} {\bibfnamefont {I.}~\bibnamefont {Bloch}},\ }\bibfield  {title} {\bibinfo {title} {{Probing the relaxation towards equilibrium in an isolated strongly correlated one-dimensional Bose gas}},\ }\href {https://doi.org/10.1038/nphys2232} {\bibfield  {journal} {\bibinfo  {journal} {Nat. Phys.}\ }\textbf {\bibinfo {volume} {8}},\ \bibinfo {pages} {325} (\bibinfo {year} {2012})}\BibitemShut {NoStop}%
\bibitem [{\citenamefont {Kaufman}\ \emph {et~al.}(2016)\citenamefont {Kaufman}, \citenamefont {Tai}, \citenamefont {Lukin}, \citenamefont {Rispoli}, \citenamefont {Schittko}, \citenamefont {Preiss},\ and\ \citenamefont {Greiner}}]{Kaufman2016}%
    \BibitemOpen
    \bibfield  {author} {\bibinfo {author} {\bibfnamefont {A.~M.}\ \bibnamefont {Kaufman}}, \bibinfo {author} {\bibfnamefont {M.~E.}\ \bibnamefont {Tai}}, \bibinfo {author} {\bibfnamefont {A.}~\bibnamefont {Lukin}}, \bibinfo {author} {\bibfnamefont {M.}~\bibnamefont {Rispoli}}, \bibinfo {author} {\bibfnamefont {R.}~\bibnamefont {Schittko}}, \bibinfo {author} {\bibfnamefont {P.~M.}\ \bibnamefont {Preiss}},\ and\ \bibinfo {author} {\bibfnamefont {M.}~\bibnamefont {Greiner}},\ }\bibfield  {title} {\bibinfo {title} {Quantum thermalization through entanglement in an isolated many-body system},\ }\href {https://doi.org/10.1126/science.aaf6725} {\bibfield  {journal} {\bibinfo  {journal} {Science}\ }\textbf {\bibinfo {volume} {353}},\ \bibinfo {pages} {794} (\bibinfo {year} {2016})}\BibitemShut {NoStop}%
\bibitem [{\citenamefont {Wang}\ \emph {et~al.}(2022)\citenamefont {Wang}, \citenamefont {Tao}, \citenamefont {Pan}, \citenamefont {Chen}, \citenamefont {Chen}, \citenamefont {Sun}, \citenamefont {Xu}, \citenamefont {Xu}, \citenamefont {Han}, \citenamefont {Li},\ and\ \citenamefont {Guo}}]{Wang2022}%
    \BibitemOpen
    \bibfield  {author} {\bibinfo {author} {\bibfnamefont {Q.-Q.}\ \bibnamefont {Wang}}, \bibinfo {author} {\bibfnamefont {S.-J.}\ \bibnamefont {Tao}}, \bibinfo {author} {\bibfnamefont {W.-W.}\ \bibnamefont {Pan}}, \bibinfo {author} {\bibfnamefont {Z.}~\bibnamefont {Chen}}, \bibinfo {author} {\bibfnamefont {G.}~\bibnamefont {Chen}}, \bibinfo {author} {\bibfnamefont {K.}~\bibnamefont {Sun}}, \bibinfo {author} {\bibfnamefont {J.-S.}\ \bibnamefont {Xu}}, \bibinfo {author} {\bibfnamefont {X.-Y.}\ \bibnamefont {Xu}}, \bibinfo {author} {\bibfnamefont {Y.-J.}\ \bibnamefont {Han}}, \bibinfo {author} {\bibfnamefont {C.-F.}\ \bibnamefont {Li}},\ and\ \bibinfo {author} {\bibfnamefont {G.-C.}\ \bibnamefont {Guo}},\ }\bibfield  {title} {\bibinfo {title} {{Experimental verification of generalized eigenstate thermalization hypothesis in an integrable system}},\ }\href {https://doi.org/10.1038/s41377-022-00887-5} {\bibfield  {journal} {\bibinfo  {journal} {Light. Sci. Appl.}\ }\textbf {\bibinfo {volume} {11}},\ \bibinfo {pages} {194} (\bibinfo {year} {2022})}\BibitemShut {NoStop}%
\bibitem [{\citenamefont {Navon}\ \emph {et~al.}(2011)\citenamefont {Navon}, \citenamefont {Piatecki}, \citenamefont {G\"unter}, \citenamefont {Rem}, \citenamefont {Nguyen}, \citenamefont {Chevy}, \citenamefont {Krauth},\ and\ \citenamefont {Salomon}}]{PhysRevLett.107.135301}%
    \BibitemOpen
    \bibfield  {author} {\bibinfo {author} {\bibfnamefont {N.}~\bibnamefont {Navon}}, \bibinfo {author} {\bibfnamefont {S.}~\bibnamefont {Piatecki}}, \bibinfo {author} {\bibfnamefont {K.}~\bibnamefont {G\"unter}}, \bibinfo {author} {\bibfnamefont {B.}~\bibnamefont {Rem}}, \bibinfo {author} {\bibfnamefont {T.~C.}\ \bibnamefont {Nguyen}}, \bibinfo {author} {\bibfnamefont {F.}~\bibnamefont {Chevy}}, \bibinfo {author} {\bibfnamefont {W.}~\bibnamefont {Krauth}},\ and\ \bibinfo {author} {\bibfnamefont {C.}~\bibnamefont {Salomon}},\ }\bibfield  {title} {\bibinfo {title} {{Dynamics and Thermodynamics of the Low-Temperature Strongly Interacting Bose Gas}},\ }\href {https://doi.org/10.1103/PhysRevLett.107.135301} {\bibfield  {journal} {\bibinfo  {journal} {Phys. Rev. Lett.}\ }\textbf {\bibinfo {volume} {107}},\ \bibinfo {pages} {135301} (\bibinfo {year} {2011})}\BibitemShut {NoStop}%
\bibitem [{\citenamefont {Ashida}\ \emph {et~al.}(2018)\citenamefont {Ashida}, \citenamefont {Saito},\ and\ \citenamefont {Ueda}}]{PhysRevLett.121.170402}%
    \BibitemOpen
    \bibfield  {author} {\bibinfo {author} {\bibfnamefont {Y.}~\bibnamefont {Ashida}}, \bibinfo {author} {\bibfnamefont {K.}~\bibnamefont {Saito}},\ and\ \bibinfo {author} {\bibfnamefont {M.}~\bibnamefont {Ueda}},\ }\bibfield  {title} {\bibinfo {title} {{Thermalization and Heating Dynamics in Open Generic Many-Body Systems}},\ }\href {https://doi.org/10.1103/PhysRevLett.121.170402} {\bibfield  {journal} {\bibinfo  {journal} {Phys. Rev. Lett.}\ }\textbf {\bibinfo {volume} {121}},\ \bibinfo {pages} {170402} (\bibinfo {year} {2018})}\BibitemShut {NoStop}%
\bibitem [{\citenamefont {Steinigeweg}\ \emph {et~al.}(2014)\citenamefont {Steinigeweg}, \citenamefont {Khodja}, \citenamefont {Niemeyer}, \citenamefont {Gogolin},\ and\ \citenamefont {Gemmer}}]{PhysRevLett.112.130403}%
    \BibitemOpen
    \bibfield  {author} {\bibinfo {author} {\bibfnamefont {R.}~\bibnamefont {Steinigeweg}}, \bibinfo {author} {\bibfnamefont {A.}~\bibnamefont {Khodja}}, \bibinfo {author} {\bibfnamefont {H.}~\bibnamefont {Niemeyer}}, \bibinfo {author} {\bibfnamefont {C.}~\bibnamefont {Gogolin}},\ and\ \bibinfo {author} {\bibfnamefont {J.}~\bibnamefont {Gemmer}},\ }\bibfield  {title} {\bibinfo {title} {{Pushing the Limits of the Eigenstate Thermalization Hypothesis towards Mesoscopic Quantum Systems}},\ }\href {https://doi.org/10.1103/PhysRevLett.112.130403} {\bibfield  {journal} {\bibinfo  {journal} {Phys. Rev. Lett.}\ }\textbf {\bibinfo {volume} {112}},\ \bibinfo {pages} {130403} (\bibinfo {year} {2014})}\BibitemShut {NoStop}%
\bibitem [{\citenamefont {Neill}\ \emph {et~al.}(2016)\citenamefont {Neill}, \citenamefont {Roushan}, \citenamefont {Fang}, \citenamefont {Chen}, \citenamefont {Kolodrubetz}, \citenamefont {Chen}, \citenamefont {Megrant}, \citenamefont {Barends}, \citenamefont {Campbell}, \citenamefont {Chiaro}, \citenamefont {Dunsworth}, \citenamefont {Jeffrey}, \citenamefont {Kelly}, \citenamefont {Mutus}, \citenamefont {O'Malley}, \citenamefont {Quintana}, \citenamefont {Sank}, \citenamefont {Vainsencher}, \citenamefont {Wenner}, \citenamefont {White}, \citenamefont {Polkovnikov},\ and\ \citenamefont {Martinis}}]{Neill2016}%
    \BibitemOpen
    \bibfield  {author} {\bibinfo {author} {\bibfnamefont {C.}~\bibnamefont {Neill}}, \bibinfo {author} {\bibfnamefont {P.}~\bibnamefont {Roushan}}, \bibinfo {author} {\bibfnamefont {M.}~\bibnamefont {Fang}}, \bibinfo {author} {\bibfnamefont {Y.}~\bibnamefont {Chen}}, \bibinfo {author} {\bibfnamefont {M.}~\bibnamefont {Kolodrubetz}}, \bibinfo {author} {\bibfnamefont {Z.}~\bibnamefont {Chen}}, \bibinfo {author} {\bibfnamefont {A.}~\bibnamefont {Megrant}}, \bibinfo {author} {\bibfnamefont {R.}~\bibnamefont {Barends}}, \bibinfo {author} {\bibfnamefont {B.}~\bibnamefont {Campbell}}, \bibinfo {author} {\bibfnamefont {B.}~\bibnamefont {Chiaro}}, \bibinfo {author} {\bibfnamefont {A.}~\bibnamefont {Dunsworth}}, \bibinfo {author} {\bibfnamefont {E.}~\bibnamefont {Jeffrey}}, \bibinfo {author} {\bibfnamefont {J.}~\bibnamefont {Kelly}}, \bibinfo {author} {\bibfnamefont {J.}~\bibnamefont {Mutus}}, \bibinfo {author} {\bibfnamefont {P.~J.~J.}\ \bibnamefont {O'Malley}}, \bibinfo {author} {\bibfnamefont {C.}~\bibnamefont {Quintana}}, \bibinfo {author} {\bibfnamefont {D.}~\bibnamefont {Sank}}, \bibinfo {author} {\bibfnamefont {A.}~\bibnamefont {Vainsencher}}, \bibinfo {author} {\bibfnamefont {J.}~\bibnamefont {Wenner}}, \bibinfo {author} {\bibfnamefont {T.~C.}\ \bibnamefont {White}}, \bibinfo {author} {\bibfnamefont {A.}~\bibnamefont {Polkovnikov}},\ and\ \bibinfo {author} {\bibfnamefont {J.~M.}\ \bibnamefont {Martinis}},\ }\bibfield  {title} {\bibinfo {title} {{Ergodic dynamics and thermalization in an isolated quantum system}},\ }\href {https://doi.org/10.1038/nphys3830} {\bibfield  {journal} {\bibinfo  {journal} {Nat. Phys.}\ }\textbf {\bibinfo {volume} {12}},\ \bibinfo {pages} {1037} (\bibinfo {year} {2016})}\BibitemShut {NoStop}%
\bibitem [{\citenamefont {Tang}\ \emph {et~al.}(2018)\citenamefont {Tang}, \citenamefont {Kao}, \citenamefont {Li}, \citenamefont {Seo}, \citenamefont {Mallayya}, \citenamefont {Rigol}, \citenamefont {Gopalakrishnan},\ and\ \citenamefont {Lev}}]{PhysRevX.8.021030}%
    \BibitemOpen
    \bibfield  {author} {\bibinfo {author} {\bibfnamefont {Y.}~\bibnamefont {Tang}}, \bibinfo {author} {\bibfnamefont {W.}~\bibnamefont {Kao}}, \bibinfo {author} {\bibfnamefont {K.-Y.}\ \bibnamefont {Li}}, \bibinfo {author} {\bibfnamefont {S.}~\bibnamefont {Seo}}, \bibinfo {author} {\bibfnamefont {K.}~\bibnamefont {Mallayya}}, \bibinfo {author} {\bibfnamefont {M.}~\bibnamefont {Rigol}}, \bibinfo {author} {\bibfnamefont {S.}~\bibnamefont {Gopalakrishnan}},\ and\ \bibinfo {author} {\bibfnamefont {B.~L.}\ \bibnamefont {Lev}},\ }\bibfield  {title} {\bibinfo {title} {Thermalization near integrability in a dipolar quantum newton's cradle},\ }\href {https://doi.org/10.1103/PhysRevX.8.021030} {\bibfield  {journal} {\bibinfo  {journal} {Phys. Rev. X}\ }\textbf {\bibinfo {volume} {8}},\ \bibinfo {pages} {021030} (\bibinfo {year} {2018})}\BibitemShut {NoStop}%
\bibitem [{\citenamefont {Sheng}\ \emph {et~al.}(2023)\citenamefont {Sheng}, \citenamefont {Yang},\ and\ \citenamefont {Wu}}]{SHENG202375}%
    \BibitemOpen
    \bibfield  {author} {\bibinfo {author} {\bibfnamefont {J.}~\bibnamefont {Sheng}}, \bibinfo {author} {\bibfnamefont {C.}~\bibnamefont {Yang}},\ and\ \bibinfo {author} {\bibfnamefont {H.}~\bibnamefont {Wu}},\ }\bibfield  {title} {\bibinfo {title} {Nonequilibrium thermodynamics in cavity optomechanics},\ }\href {https://doi.org/https://doi.org/10.1016/j.fmre.2022.09.005} {\bibfield  {journal} {\bibinfo  {journal} {Fundam. Res.}\ }\textbf {\bibinfo {volume} {3}},\ \bibinfo {pages} {75} (\bibinfo {year} {2023})}\BibitemShut {NoStop}%
\bibitem [{\citenamefont {Polkovnikov}\ \emph {et~al.}(2011)\citenamefont {Polkovnikov}, \citenamefont {Sengupta}, \citenamefont {Silva},\ and\ \citenamefont {Vengalattore}}]{RevModPhys.83.863}%
    \BibitemOpen
    \bibfield  {author} {\bibinfo {author} {\bibfnamefont {A.}~\bibnamefont {Polkovnikov}}, \bibinfo {author} {\bibfnamefont {K.}~\bibnamefont {Sengupta}}, \bibinfo {author} {\bibfnamefont {A.}~\bibnamefont {Silva}},\ and\ \bibinfo {author} {\bibfnamefont {M.}~\bibnamefont {Vengalattore}},\ }\bibfield  {title} {\bibinfo {title} {{Colloquium: Nonequilibrium dynamics of closed interacting quantum systems}},\ }\href {https://doi.org/10.1103/RevModPhys.83.863} {\bibfield  {journal} {\bibinfo  {journal} {Rev. Mod. Phys.}\ }\textbf {\bibinfo {volume} {83}},\ \bibinfo {pages} {863} (\bibinfo {year} {2011})}\BibitemShut {NoStop}%
\bibitem [{\citenamefont {Ji}\ \emph {et~al.}(2014)\citenamefont {Ji}, \citenamefont {Zhang}, \citenamefont {Zhang}, \citenamefont {Du}, \citenamefont {Zheng}, \citenamefont {Deng}, \citenamefont {Zhai}, \citenamefont {Chen},\ and\ \citenamefont {Pan}}]{ji2014experimental}%
    \BibitemOpen
    \bibfield  {author} {\bibinfo {author} {\bibfnamefont {S.-C.}\ \bibnamefont {Ji}}, \bibinfo {author} {\bibfnamefont {J.-Y.}\ \bibnamefont {Zhang}}, \bibinfo {author} {\bibfnamefont {L.}~\bibnamefont {Zhang}}, \bibinfo {author} {\bibfnamefont {Z.-D.}\ \bibnamefont {Du}}, \bibinfo {author} {\bibfnamefont {W.}~\bibnamefont {Zheng}}, \bibinfo {author} {\bibfnamefont {Y.-J.}\ \bibnamefont {Deng}}, \bibinfo {author} {\bibfnamefont {H.}~\bibnamefont {Zhai}}, \bibinfo {author} {\bibfnamefont {S.}~\bibnamefont {Chen}},\ and\ \bibinfo {author} {\bibfnamefont {J.-W.}\ \bibnamefont {Pan}},\ }\bibfield  {title} {\bibinfo {title} {Experimental determination of the finite-temperature phase diagram of a spin--orbit coupled bose gas},\ }\href {https://doi.org/10.1038/nphys2905} {\bibfield  {journal} {\bibinfo  {journal} {Nat. phys.}\ }\textbf {\bibinfo {volume} {10}},\ \bibinfo {pages} {314} (\bibinfo {year} {2014})}\BibitemShut {NoStop}%
\bibitem [{\citenamefont {Eigen}\ \emph {et~al.}(2018)\citenamefont {Eigen}, \citenamefont {Glidden}, \citenamefont {Lopes}, \citenamefont {Cornell}, \citenamefont {Smith},\ and\ \citenamefont {Hadzibabic}}]{eigen2018universal}%
    \BibitemOpen
    \bibfield  {author} {\bibinfo {author} {\bibfnamefont {C.}~\bibnamefont {Eigen}}, \bibinfo {author} {\bibfnamefont {J.~A.}\ \bibnamefont {Glidden}}, \bibinfo {author} {\bibfnamefont {R.}~\bibnamefont {Lopes}}, \bibinfo {author} {\bibfnamefont {E.~A.}\ \bibnamefont {Cornell}}, \bibinfo {author} {\bibfnamefont {R.~P.}\ \bibnamefont {Smith}},\ and\ \bibinfo {author} {\bibfnamefont {Z.}~\bibnamefont {Hadzibabic}},\ }\bibfield  {title} {\bibinfo {title} {Universal prethermal dynamics of bose gases quenched to unitarity},\ }\href {https://doi.org/10.1038/nphys2905} {\bibfield  {journal} {\bibinfo  {journal} {Nature}\ }\textbf {\bibinfo {volume} {563}},\ \bibinfo {pages} {221} (\bibinfo {year} {2018})}\BibitemShut {NoStop}%
\bibitem [{\citenamefont {Bloch}\ \emph {et~al.}(2012)\citenamefont {Bloch}, \citenamefont {Dalibard},\ and\ \citenamefont {Nascimb{\`{e}}ne}}]{Bloch2012}%
    \BibitemOpen
    \bibfield  {author} {\bibinfo {author} {\bibfnamefont {I.}~\bibnamefont {Bloch}}, \bibinfo {author} {\bibfnamefont {J.}~\bibnamefont {Dalibard}},\ and\ \bibinfo {author} {\bibfnamefont {S.}~\bibnamefont {Nascimb{\`{e}}ne}},\ }\bibfield  {title} {\bibinfo {title} {{Quantum simulations with ultracold quantum gases}},\ }\href {https://doi.org/10.1038/nphys2259} {\bibfield  {journal} {\bibinfo  {journal} {Nat. Phys.}\ }\textbf {\bibinfo {volume} {8}},\ \bibinfo {pages} {267} (\bibinfo {year} {2012})}\BibitemShut {NoStop}%
\bibitem [{\citenamefont {Gross}\ and\ \citenamefont {Bloch}(2017)}]{science.aal3837}%
    \BibitemOpen
    \bibfield  {author} {\bibinfo {author} {\bibfnamefont {C.}~\bibnamefont {Gross}}\ and\ \bibinfo {author} {\bibfnamefont {I.}~\bibnamefont {Bloch}},\ }\bibfield  {title} {\bibinfo {title} {Quantum simulations with ultracold atoms in optical lattices},\ }\href {https://doi.org/10.1126/science.aal3837} {\bibfield  {journal} {\bibinfo  {journal} {Science}\ }\textbf {\bibinfo {volume} {357}},\ \bibinfo {pages} {995} (\bibinfo {year} {2017})}\BibitemShut {NoStop}%
\bibitem [{\citenamefont {Greiner}\ \emph {et~al.}(2002)\citenamefont {Greiner}, \citenamefont {Mandel}, \citenamefont {Esslinger}, \citenamefont {H{\"{a}}nsch},\ and\ \citenamefont {Bloch}}]{Greiner2002}%
    \BibitemOpen
    \bibfield  {author} {\bibinfo {author} {\bibfnamefont {M.}~\bibnamefont {Greiner}}, \bibinfo {author} {\bibfnamefont {O.}~\bibnamefont {Mandel}}, \bibinfo {author} {\bibfnamefont {T.}~\bibnamefont {Esslinger}}, \bibinfo {author} {\bibfnamefont {T.~W.}\ \bibnamefont {H{\"{a}}nsch}},\ and\ \bibinfo {author} {\bibfnamefont {I.}~\bibnamefont {Bloch}},\ }\bibfield  {title} {\bibinfo {title} {{Quantum phase transition from a superfluid to a Mott insulator in a gas of ultracold atoms}},\ }\href {https://doi.org/10.1038/415039a} {\bibfield  {journal} {\bibinfo  {journal} {Nature}\ }\textbf {\bibinfo {volume} {415}},\ \bibinfo {pages} {39} (\bibinfo {year} {2002})}\BibitemShut {NoStop}%
\bibitem [{\citenamefont {Pupillo}\ \emph {et~al.}(2006)\citenamefont {Pupillo}, \citenamefont {Williams},\ and\ \citenamefont {Prokof'ev}}]{PhysRevA.73.013408}%
    \BibitemOpen
    \bibfield  {author} {\bibinfo {author} {\bibfnamefont {G.}~\bibnamefont {Pupillo}}, \bibinfo {author} {\bibfnamefont {C.~J.}\ \bibnamefont {Williams}},\ and\ \bibinfo {author} {\bibfnamefont {N.~V.}\ \bibnamefont {Prokof'ev}},\ }\bibfield  {title} {\bibinfo {title} {Effects of finite temperature on the mott-insulator state},\ }\href {https://doi.org/10.1103/PhysRevA.73.013408} {\bibfield  {journal} {\bibinfo  {journal} {Phys. Rev. A}\ }\textbf {\bibinfo {volume} {73}},\ \bibinfo {pages} {013408} (\bibinfo {year} {2006})}\BibitemShut {NoStop}%
\bibitem [{\citenamefont {Polkovnikov}(2005)}]{PhysRevB.72.161201}%
    \BibitemOpen
    \bibfield  {author} {\bibinfo {author} {\bibfnamefont {A.}~\bibnamefont {Polkovnikov}},\ }\bibfield  {title} {\bibinfo {title} {Universal adiabatic dynamics in the vicinity of a quantum critical point},\ }\href {https://doi.org/10.1103/PhysRevB.72.161201} {\bibfield  {journal} {\bibinfo  {journal} {Phys. Rev. B}\ }\textbf {\bibinfo {volume} {72}},\ \bibinfo {pages} {161201} (\bibinfo {year} {2005})}\BibitemShut {NoStop}%
\bibitem [{\citenamefont {Dziarmaga}(2010)}]{Dziarmaga2010}%
    \BibitemOpen
    \bibfield  {author} {\bibinfo {author} {\bibfnamefont {J.}~\bibnamefont {Dziarmaga}},\ }\bibfield  {title} {\bibinfo {title} {Dynamics of a quantum phase transition and relaxation to a steady state},\ }\href {https://doi.org/10.1080/00018732.2010.514702} {\bibfield  {journal} {\bibinfo  {journal} {Adv. Phys.}\ }\textbf {\bibinfo {volume} {59}},\ \bibinfo {pages} {1063} (\bibinfo {year} {2010})}\BibitemShut {NoStop}%
\bibitem [{\citenamefont {del Campo}\ and\ \citenamefont {Zurek}(2014)}]{Campo2014}%
    \BibitemOpen
    \bibfield  {author} {\bibinfo {author} {\bibfnamefont {A.}~\bibnamefont {del Campo}}\ and\ \bibinfo {author} {\bibfnamefont {W.~H.}\ \bibnamefont {Zurek}},\ }\bibfield  {title} {\bibinfo {title} {Universality of phase transition dynamics: Topological defects from symmetry breaking},\ }\href {https://doi.org/10.1142/S0217751X1430018X} {\bibfield  {journal} {\bibinfo  {journal} {Int. J. Mod. Phys. A}\ }\textbf {\bibinfo {volume} {29}},\ \bibinfo {pages} {1430018} (\bibinfo {year} {2014})}\BibitemShut {NoStop}%
\bibitem [{\citenamefont {Huang}\ \emph {et~al.}(2021)\citenamefont {Huang}, \citenamefont {Yao}, \citenamefont {Liang}, \citenamefont {Wang}, \citenamefont {Zheng}, \citenamefont {Li}, \citenamefont {Xiong}, \citenamefont {Zhou}, \citenamefont {Chen}, \citenamefont {Chen},\ and\ \citenamefont {Hu}}]{PhysRevLett.127.200601}%
    \BibitemOpen
    \bibfield  {author} {\bibinfo {author} {\bibfnamefont {Q.}~\bibnamefont {Huang}}, \bibinfo {author} {\bibfnamefont {R.}~\bibnamefont {Yao}}, \bibinfo {author} {\bibfnamefont {L.}~\bibnamefont {Liang}}, \bibinfo {author} {\bibfnamefont {S.}~\bibnamefont {Wang}}, \bibinfo {author} {\bibfnamefont {Q.}~\bibnamefont {Zheng}}, \bibinfo {author} {\bibfnamefont {D.}~\bibnamefont {Li}}, \bibinfo {author} {\bibfnamefont {W.}~\bibnamefont {Xiong}}, \bibinfo {author} {\bibfnamefont {X.}~\bibnamefont {Zhou}}, \bibinfo {author} {\bibfnamefont {W.}~\bibnamefont {Chen}}, \bibinfo {author} {\bibfnamefont {X.}~\bibnamefont {Chen}},\ and\ \bibinfo {author} {\bibfnamefont {J.}~\bibnamefont {Hu}},\ }\bibfield  {title} {\bibinfo {title} {Observation of many-body quantum phase transitions beyond the kibble-zurek mechanism},\ }\href {https://doi.org/10.1103/PhysRevLett.127.200601} {\bibfield  {journal} {\bibinfo  {journal} {Phys. Rev. Lett.}\ }\textbf {\bibinfo {volume} {127}},\ \bibinfo {pages} {200601} (\bibinfo {year} {2021})}\BibitemShut {NoStop}%
\bibitem [{\citenamefont {{K{\"o}hl}}\ \emph {et~al.}(2005)\citenamefont {{K{\"o}hl}}, \citenamefont {{Moritz}}, \citenamefont {{St{\"o}ferle}}, \citenamefont {{G{\"u}nter}},\ and\ \citenamefont {{Esslinger}}}]{PhysRevLett.94.080403}%
    \BibitemOpen
    \bibfield  {author} {\bibinfo {author} {\bibfnamefont {M.}~\bibnamefont {{K{\"o}hl}}}, \bibinfo {author} {\bibfnamefont {H.}~\bibnamefont {{Moritz}}}, \bibinfo {author} {\bibfnamefont {T.}~\bibnamefont {{St{\"o}ferle}}}, \bibinfo {author} {\bibfnamefont {K.}~\bibnamefont {{G{\"u}nter}}},\ and\ \bibinfo {author} {\bibfnamefont {T.}~\bibnamefont {{Esslinger}}},\ }\bibfield  {title} {\bibinfo {title} {Fermionic atoms in a three dimensional optical lattice: Observing fermi surfaces, dynamics, and interactions},\ }\href {https://doi.org/10.1103/PhysRevLett.94.080403} {\bibfield  {journal} {\bibinfo  {journal} {Phys. Rev. Lett.}\ }\textbf {\bibinfo {volume} {94}},\ \bibinfo {pages} {080403} (\bibinfo {year} {2005})}\BibitemShut {NoStop}%
\bibitem [{\citenamefont {Liang}\ \emph {et~al.}(2022)\citenamefont {Liang}, \citenamefont {Zheng}, \citenamefont {Yao}, \citenamefont {Zheng}, \citenamefont {Yao}, \citenamefont {Zhou}, \citenamefont {Huang}, \citenamefont {Zhang}, \citenamefont {Ye}, \citenamefont {Zhou}, \citenamefont {Chen}, \citenamefont {Chen}, \citenamefont {Zhai},\ and\ \citenamefont {Hu}}]{LIANG20222550}%
    \BibitemOpen
    \bibfield  {author} {\bibinfo {author} {\bibfnamefont {L.}~\bibnamefont {Liang}}, \bibinfo {author} {\bibfnamefont {W.}~\bibnamefont {Zheng}}, \bibinfo {author} {\bibfnamefont {R.}~\bibnamefont {Yao}}, \bibinfo {author} {\bibfnamefont {Q.}~\bibnamefont {Zheng}}, \bibinfo {author} {\bibfnamefont {Z.}~\bibnamefont {Yao}}, \bibinfo {author} {\bibfnamefont {T.-G.}\ \bibnamefont {Zhou}}, \bibinfo {author} {\bibfnamefont {Q.}~\bibnamefont {Huang}}, \bibinfo {author} {\bibfnamefont {Z.}~\bibnamefont {Zhang}}, \bibinfo {author} {\bibfnamefont {J.}~\bibnamefont {Ye}}, \bibinfo {author} {\bibfnamefont {X.}~\bibnamefont {Zhou}}, \bibinfo {author} {\bibfnamefont {X.}~\bibnamefont {Chen}}, \bibinfo {author} {\bibfnamefont {W.}~\bibnamefont {Chen}}, \bibinfo {author} {\bibfnamefont {H.}~\bibnamefont {Zhai}},\ and\ \bibinfo {author} {\bibfnamefont {J.}~\bibnamefont {Hu}},\ }\bibfield  {title} {\bibinfo {title} {Probing quantum many-body correlations by universal ramping dynamics},\ }\href {https://doi.org/https://doi.org/10.1016/j.scib.2022.12.005} {\bibfield  {journal} {\bibinfo  {journal} {Sci. Bull.}\ }\textbf {\bibinfo {volume} {67}},\ \bibinfo {pages} {2550} (\bibinfo {year} {2022})}\BibitemShut {NoStop}%
\bibitem [{\citenamefont {Zheng}\ \emph {et~al.}(2023)\citenamefont {Zheng}, \citenamefont {Wang}, \citenamefont {Liang}, \citenamefont {Huang}, \citenamefont {Wang}, \citenamefont {Xiong}, \citenamefont {Zhou}, \citenamefont {Chen}, \citenamefont {Chen},\ and\ \citenamefont {Hu}}]{PhysRevResearch.5.013136}%
    \BibitemOpen
    \bibfield  {author} {\bibinfo {author} {\bibfnamefont {Q.}~\bibnamefont {Zheng}}, \bibinfo {author} {\bibfnamefont {Y.}~\bibnamefont {Wang}}, \bibinfo {author} {\bibfnamefont {L.}~\bibnamefont {Liang}}, \bibinfo {author} {\bibfnamefont {Q.}~\bibnamefont {Huang}}, \bibinfo {author} {\bibfnamefont {S.}~\bibnamefont {Wang}}, \bibinfo {author} {\bibfnamefont {W.}~\bibnamefont {Xiong}}, \bibinfo {author} {\bibfnamefont {X.}~\bibnamefont {Zhou}}, \bibinfo {author} {\bibfnamefont {W.}~\bibnamefont {Chen}}, \bibinfo {author} {\bibfnamefont {X.}~\bibnamefont {Chen}},\ and\ \bibinfo {author} {\bibfnamefont {J.}~\bibnamefont {Hu}},\ }\bibfield  {title} {\bibinfo {title} {Dimensional crossover of quantum critical dynamics in many-body phase transitions},\ }\href {https://doi.org/10.1103/PhysRevResearch.5.013136} {\bibfield  {journal} {\bibinfo  {journal} {Phys. Rev. Res.}\ }\textbf {\bibinfo {volume} {5}},\ \bibinfo {pages} {013136} (\bibinfo {year} {2023})}\BibitemShut {NoStop}%
\bibitem [{\citenamefont {Sherson}\ \emph {et~al.}(2010)\citenamefont {Sherson}, \citenamefont {Weitenberg}, \citenamefont {Endres}, \citenamefont {Cheneau}, \citenamefont {Bloch},\ and\ \citenamefont {Kuhr}}]{Sherson2010}%
    \BibitemOpen
    \bibfield  {author} {\bibinfo {author} {\bibfnamefont {J.~F.}\ \bibnamefont {Sherson}}, \bibinfo {author} {\bibfnamefont {C.}~\bibnamefont {Weitenberg}}, \bibinfo {author} {\bibfnamefont {M.}~\bibnamefont {Endres}}, \bibinfo {author} {\bibfnamefont {M.}~\bibnamefont {Cheneau}}, \bibinfo {author} {\bibfnamefont {I.}~\bibnamefont {Bloch}},\ and\ \bibinfo {author} {\bibfnamefont {S.}~\bibnamefont {Kuhr}},\ }\bibfield  {title} {\bibinfo {title} {{Single-atom-resolved fluorescence imaging of an atomic Mott insulator}},\ }\href {https://doi.org/10.1038/nature09378} {\bibfield  {journal} {\bibinfo  {journal} {Nature}\ }\textbf {\bibinfo {volume} {467}},\ \bibinfo {pages} {68} (\bibinfo {year} {2010})}\BibitemShut {NoStop}%
\bibitem [{\citenamefont {Fisher}\ \emph {et~al.}(1989)\citenamefont {Fisher}, \citenamefont {Weichman}, \citenamefont {Grinstein},\ and\ \citenamefont {Fisher}}]{PhysRevB.40.546}%
    \BibitemOpen
    \bibfield  {author} {\bibinfo {author} {\bibfnamefont {M.~P.~A.}\ \bibnamefont {Fisher}}, \bibinfo {author} {\bibfnamefont {P.~B.}\ \bibnamefont {Weichman}}, \bibinfo {author} {\bibfnamefont {G.}~\bibnamefont {Grinstein}},\ and\ \bibinfo {author} {\bibfnamefont {D.~S.}\ \bibnamefont {Fisher}},\ }\bibfield  {title} {\bibinfo {title} {Boson localization and the superfluid-insulator transition},\ }\href {https://doi.org/10.1103/PhysRevB.40.546} {\bibfield  {journal} {\bibinfo  {journal} {Phys. Rev. B}\ }\textbf {\bibinfo {volume} {40}},\ \bibinfo {pages} {546} (\bibinfo {year} {1989})}\BibitemShut {NoStop}%
\bibitem [{\citenamefont {Gerbier}\ \emph {et~al.}(2005)\citenamefont {Gerbier}, \citenamefont {Widera}, \citenamefont {F\"olling}, \citenamefont {Mandel}, \citenamefont {Gericke},\ and\ \citenamefont {Bloch}}]{PhysRevA.72.053606}%
    \BibitemOpen
    \bibfield  {author} {\bibinfo {author} {\bibfnamefont {F.}~\bibnamefont {Gerbier}}, \bibinfo {author} {\bibfnamefont {A.}~\bibnamefont {Widera}}, \bibinfo {author} {\bibfnamefont {S.}~\bibnamefont {F\"olling}}, \bibinfo {author} {\bibfnamefont {O.}~\bibnamefont {Mandel}}, \bibinfo {author} {\bibfnamefont {T.}~\bibnamefont {Gericke}},\ and\ \bibinfo {author} {\bibfnamefont {I.}~\bibnamefont {Bloch}},\ }\bibfield  {title} {\bibinfo {title} {Interference pattern and visibility of a mott insulator},\ }\href {https://doi.org/10.1103/PhysRevA.72.053606} {\bibfield  {journal} {\bibinfo  {journal} {Phys. Rev. A}\ }\textbf {\bibinfo {volume} {72}},\ \bibinfo {pages} {053606} (\bibinfo {year} {2005})}\BibitemShut {NoStop}%
\bibitem [{\citenamefont {Liang}\ \emph {et~al.}(2023)\citenamefont {Liang}, \citenamefont {Wang}, \citenamefont {Huang}, \citenamefont {Zheng}, \citenamefont {Chen},\ and\ \citenamefont {Hu}}]{Liang23}%
    \BibitemOpen
    \bibfield  {author} {\bibinfo {author} {\bibfnamefont {L.}~\bibnamefont {Liang}}, \bibinfo {author} {\bibfnamefont {Y.}~\bibnamefont {Wang}}, \bibinfo {author} {\bibfnamefont {Q.}~\bibnamefont {Huang}}, \bibinfo {author} {\bibfnamefont {Q.}~\bibnamefont {Zheng}}, \bibinfo {author} {\bibfnamefont {X.}~\bibnamefont {Chen}},\ and\ \bibinfo {author} {\bibfnamefont {J.}~\bibnamefont {Hu}},\ }\bibfield  {title} {\bibinfo {title} {Probing quantum phase transition point by tuning an external anti trap},\ }\href {https://doi.org/10.1364/OE.487196} {\bibfield  {journal} {\bibinfo  {journal} {Opt. Express}\ }\textbf {\bibinfo {volume} {31}},\ \bibinfo {pages} {16743} (\bibinfo {year} {2023})}\BibitemShut {NoStop}%
\bibitem [{\citenamefont {Carcy}\ \emph {et~al.}(2021)\citenamefont {Carcy}, \citenamefont {Herc{\'e}}, \citenamefont {Tenart}, \citenamefont {Roscilde},\ and\ \citenamefont {Cl{\'e}ment}}]{carcy2021certifying}%
    \BibitemOpen
    \bibfield  {author} {\bibinfo {author} {\bibfnamefont {C.}~\bibnamefont {Carcy}}, \bibinfo {author} {\bibfnamefont {G.}~\bibnamefont {Herc{\'e}}}, \bibinfo {author} {\bibfnamefont {A.}~\bibnamefont {Tenart}}, \bibinfo {author} {\bibfnamefont {T.}~\bibnamefont {Roscilde}},\ and\ \bibinfo {author} {\bibfnamefont {D.}~\bibnamefont {Cl{\'e}ment}},\ }\bibfield  {title} {\bibinfo {title} {Certifying the adiabatic preparation of ultracold lattice bosons in the vicinity of the mott transition},\ }\href {https://doi.org/10.1103/PhysRevLett.126.045301} {\bibfield  {journal} {\bibinfo  {journal} {Phys. Rev. Lett.}\ }\textbf {\bibinfo {volume} {126}},\ \bibinfo {pages} {045301} (\bibinfo {year} {2021})}\BibitemShut {NoStop}%
\bibitem [{\citenamefont {Miesner}\ \emph {et~al.}(1998)\citenamefont {Miesner}, \citenamefont {Stamper-Kurn}, \citenamefont {Andrews}, \citenamefont {Durfee}, \citenamefont {Inouye},\ and\ \citenamefont {Ketterle}}]{Miesner1998}%
    \BibitemOpen
    \bibfield  {author} {\bibinfo {author} {\bibfnamefont {H.-J.}\ \bibnamefont {Miesner}}, \bibinfo {author} {\bibfnamefont {D.~M.}\ \bibnamefont {Stamper-Kurn}}, \bibinfo {author} {\bibfnamefont {M.~R.}\ \bibnamefont {Andrews}}, \bibinfo {author} {\bibfnamefont {D.~S.}\ \bibnamefont {Durfee}}, \bibinfo {author} {\bibfnamefont {S.}~\bibnamefont {Inouye}},\ and\ \bibinfo {author} {\bibfnamefont {W.}~\bibnamefont {Ketterle}},\ }\bibfield  {title} {\bibinfo {title} {{Bosonic Stimulation in the Formation of a Bose-Einstein Condensate}},\ }\href {https://doi.org/10.1126/science.279.5353.1005} {\bibfield  {journal} {\bibinfo  {journal} {Science}\ }\textbf {\bibinfo {volume} {279}},\ \bibinfo {pages} {1005} (\bibinfo {year} {1998})}\BibitemShut {NoStop}%
\bibitem [{\citenamefont {Lu}\ \emph {et~al.}(2023)\citenamefont {Lu}, \citenamefont {Margalit},\ and\ \citenamefont {Ketterle}}]{Lu2023}%
    \BibitemOpen
    \bibfield  {author} {\bibinfo {author} {\bibfnamefont {Y.-K.}\ \bibnamefont {Lu}}, \bibinfo {author} {\bibfnamefont {Y.}~\bibnamefont {Margalit}},\ and\ \bibinfo {author} {\bibfnamefont {W.}~\bibnamefont {Ketterle}},\ }\bibfield  {title} {\bibinfo {title} {{Bosonic stimulation of atom–light scattering in an ultracold gas}},\ }\href {https://doi.org/10.1038/s41567-022-01846-y} {\bibfield  {journal} {\bibinfo  {journal} {Nat. Phys.}\ }\textbf {\bibinfo {volume} {19}},\ \bibinfo {pages} {210} (\bibinfo {year} {2023})}\BibitemShut {NoStop}%
\bibitem [{\citenamefont {Xing}\ \emph {et~al.}(2015)\citenamefont {Xing}, \citenamefont {Zhang}, \citenamefont {Fu}, \citenamefont {Liu}, \citenamefont {Sun}, \citenamefont {Peng}, \citenamefont {Wang}, \citenamefont {Lin}, \citenamefont {Ma}, \citenamefont {Xue}, \citenamefont {Wang},\ and\ \citenamefont {Xie}}]{10_Ying2015Science}%
    \BibitemOpen
    \bibfield  {author} {\bibinfo {author} {\bibfnamefont {Y.}~\bibnamefont {Xing}}, \bibinfo {author} {\bibfnamefont {H.-M.}\ \bibnamefont {Zhang}}, \bibinfo {author} {\bibfnamefont {H.-L.}\ \bibnamefont {Fu}}, \bibinfo {author} {\bibfnamefont {H.}~\bibnamefont {Liu}}, \bibinfo {author} {\bibfnamefont {Y.}~\bibnamefont {Sun}}, \bibinfo {author} {\bibfnamefont {J.-P.}\ \bibnamefont {Peng}}, \bibinfo {author} {\bibfnamefont {F.}~\bibnamefont {Wang}}, \bibinfo {author} {\bibfnamefont {X.}~\bibnamefont {Lin}}, \bibinfo {author} {\bibfnamefont {X.-C.}\ \bibnamefont {Ma}}, \bibinfo {author} {\bibfnamefont {Q.-K.}\ \bibnamefont {Xue}}, \bibinfo {author} {\bibfnamefont {J.}~\bibnamefont {Wang}},\ and\ \bibinfo {author} {\bibfnamefont {X.~C.}\ \bibnamefont {Xie}},\ }\bibfield  {title} {\bibinfo {title} {Quantum griffiths singularity of superconductor-metal transition in ga thin films},\ }\href {https://doi.org/10.1126/science.aaa7154} {\bibfield  {journal} {\bibinfo  {journal} {Science}\ }\textbf {\bibinfo {volume} {350}},\ \bibinfo {pages} {542} (\bibinfo {year} {2015})}\BibitemShut {NoStop}%
\bibitem [{\citenamefont {Liu}\ \emph {et~al.}(2021)\citenamefont {Liu}, \citenamefont {Qi}, \citenamefont {Fang}, \citenamefont {Sun}, \citenamefont {Liu}, \citenamefont {Liu}, \citenamefont {Qi}, \citenamefont {Xing}, \citenamefont {Liu}, \citenamefont {Lin}, \citenamefont {Wang}, \citenamefont {Xue}, \citenamefont {Xie},\ and\ \citenamefont {Wang}}]{PhysRevLett.127.137001}%
    \BibitemOpen
    \bibfield  {author} {\bibinfo {author} {\bibfnamefont {Y.}~\bibnamefont {Liu}}, \bibinfo {author} {\bibfnamefont {S.}~\bibnamefont {Qi}}, \bibinfo {author} {\bibfnamefont {J.}~\bibnamefont {Fang}}, \bibinfo {author} {\bibfnamefont {J.}~\bibnamefont {Sun}}, \bibinfo {author} {\bibfnamefont {C.}~\bibnamefont {Liu}}, \bibinfo {author} {\bibfnamefont {Y.}~\bibnamefont {Liu}}, \bibinfo {author} {\bibfnamefont {J.}~\bibnamefont {Qi}}, \bibinfo {author} {\bibfnamefont {Y.}~\bibnamefont {Xing}}, \bibinfo {author} {\bibfnamefont {H.}~\bibnamefont {Liu}}, \bibinfo {author} {\bibfnamefont {X.}~\bibnamefont {Lin}}, \bibinfo {author} {\bibfnamefont {L.}~\bibnamefont {Wang}}, \bibinfo {author} {\bibfnamefont {Q.-K.}\ \bibnamefont {Xue}}, \bibinfo {author} {\bibfnamefont {X.~C.}\ \bibnamefont {Xie}},\ and\ \bibinfo {author} {\bibfnamefont {J.}~\bibnamefont {Wang}},\ }\bibfield  {title} {\bibinfo {title} {Observation of in-plane quantum griffiths singularity in two-dimensional crystalline superconductors},\ }\href {https://doi.org/10.1103/PhysRevLett.127.137001} {\bibfield  {journal} {\bibinfo  {journal} {Phys. Rev. Lett.}\ }\textbf {\bibinfo {volume} {127}},\ \bibinfo {pages} {137001} (\bibinfo {year} {2021})}\BibitemShut {NoStop}%
\bibitem [{\citenamefont {Reiss}\ \emph {et~al.}(2021)\citenamefont {Reiss}, \citenamefont {Graf}, \citenamefont {Haghighirad}, \citenamefont {Vojta},\ and\ \citenamefont {Coldea}}]{Reiss2021}%
    \BibitemOpen
    \bibfield  {author} {\bibinfo {author} {\bibfnamefont {P.}~\bibnamefont {Reiss}}, \bibinfo {author} {\bibfnamefont {D.}~\bibnamefont {Graf}}, \bibinfo {author} {\bibfnamefont {A.~A.}\ \bibnamefont {Haghighirad}}, \bibinfo {author} {\bibfnamefont {T.}~\bibnamefont {Vojta}},\ and\ \bibinfo {author} {\bibfnamefont {A.~I.}\ \bibnamefont {Coldea}},\ }\bibfield  {title} {\bibinfo {title} {{Signatures of a Quantum Griffiths Phase Close to an Electronic Nematic Quantum Phase Transition}},\ }\href {https://doi.org/10.1103/PhysRevLett.127.246402} {\bibfield  {journal} {\bibinfo  {journal} {Phys. Rev. Lett.}\ }\textbf {\bibinfo {volume} {127}},\ \bibinfo {pages} {246402} (\bibinfo {year} {2021})}\BibitemShut {NoStop}%
\bibitem [{\citenamefont {Wang}\ \emph {et~al.}(2023)\citenamefont {Wang}, \citenamefont {Liu}, \citenamefont {Ji},\ and\ \citenamefont {Wang}}]{Wang_2024}%
    \BibitemOpen
    \bibfield  {author} {\bibinfo {author} {\bibfnamefont {Z.}~\bibnamefont {Wang}}, \bibinfo {author} {\bibfnamefont {Y.}~\bibnamefont {Liu}}, \bibinfo {author} {\bibfnamefont {C.}~\bibnamefont {Ji}},\ and\ \bibinfo {author} {\bibfnamefont {J.}~\bibnamefont {Wang}},\ }\bibfield  {title} {\bibinfo {title} {Quantum phase transitions in two-dimensional superconductors: a review on recent experimental progress},\ }\href {https://doi.org/10.1088/1361-6633/ad14f3} {\bibfield  {journal} {\bibinfo  {journal} {Rep. Prog. Phys.}\ }\textbf {\bibinfo {volume} {87}},\ \bibinfo {pages} {014502} (\bibinfo {year} {2023})}\BibitemShut {NoStop}%
\bibitem [{\citenamefont {Zurek}\ \emph {et~al.}(2005)\citenamefont {Zurek}, \citenamefont {Dorner},\ and\ \citenamefont {Zoller}}]{PhysRevLett.95.105701}%
    \BibitemOpen
    \bibfield  {author} {\bibinfo {author} {\bibfnamefont {W.~H.}\ \bibnamefont {Zurek}}, \bibinfo {author} {\bibfnamefont {U.}~\bibnamefont {Dorner}},\ and\ \bibinfo {author} {\bibfnamefont {P.}~\bibnamefont {Zoller}},\ }\bibfield  {title} {\bibinfo {title} {Dynamics of a quantum phase transition},\ }\href {https://doi.org/10.1103/PhysRevLett.95.105701} {\bibfield  {journal} {\bibinfo  {journal} {Phys. Rev. Lett.}\ }\textbf {\bibinfo {volume} {95}},\ \bibinfo {pages} {105701} (\bibinfo {year} {2005})}\BibitemShut {NoStop}%
\bibitem [{\citenamefont {Navon}\ \emph {et~al.}(2015)\citenamefont {Navon}, \citenamefont {Gaunt}, \citenamefont {Smith},\ and\ \citenamefont {Hadzibabic}}]{Navon167}%
    \BibitemOpen
    \bibfield  {author} {\bibinfo {author} {\bibfnamefont {N.}~\bibnamefont {Navon}}, \bibinfo {author} {\bibfnamefont {A.~L.}\ \bibnamefont {Gaunt}}, \bibinfo {author} {\bibfnamefont {R.~P.}\ \bibnamefont {Smith}},\ and\ \bibinfo {author} {\bibfnamefont {Z.}~\bibnamefont {Hadzibabic}},\ }\bibfield  {title} {\bibinfo {title} {Critical dynamics of spontaneous symmetry breaking in a homogeneous bose gas},\ }\href {https://doi.org/10.1126/science.1258676} {\bibfield  {journal} {\bibinfo  {journal} {Science}\ }\textbf {\bibinfo {volume} {347}},\ \bibinfo {pages} {167} (\bibinfo {year} {2015})}\BibitemShut {NoStop}%
\bibitem [{\citenamefont {Clark}\ \emph {et~al.}(2016)\citenamefont {Clark}, \citenamefont {Feng},\ and\ \citenamefont {Chin}}]{CChinKBMZ}%
    \BibitemOpen
    \bibfield  {author} {\bibinfo {author} {\bibfnamefont {L.~W.}\ \bibnamefont {Clark}}, \bibinfo {author} {\bibfnamefont {L.}~\bibnamefont {Feng}},\ and\ \bibinfo {author} {\bibfnamefont {C.}~\bibnamefont {Chin}},\ }\bibfield  {title} {\bibinfo {title} {Universal space-time scaling symmetry in the dynamics of bosons across a quantum phase transition},\ }\href {https://doi.org/10.1126/science.aaf9657} {\bibfield  {journal} {\bibinfo  {journal} {Science}\ }\textbf {\bibinfo {volume} {354}},\ \bibinfo {pages} {606–610} (\bibinfo {year} {2016})}\BibitemShut {NoStop}%
\bibitem [{\citenamefont {Ko}\ \emph {et~al.}(2019)\citenamefont {Ko}, \citenamefont {Park},\ and\ \citenamefont {Shin}}]{Ko2019}%
    \BibitemOpen
    \bibfield  {author} {\bibinfo {author} {\bibfnamefont {B.}~\bibnamefont {Ko}}, \bibinfo {author} {\bibfnamefont {J.~W.}\ \bibnamefont {Park}},\ and\ \bibinfo {author} {\bibfnamefont {Y.}~\bibnamefont {Shin}},\ }\bibfield  {title} {\bibinfo {title} {{Kibble–Zurek universality in a strongly interacting Fermi superfluid}},\ }\href {https://doi.org/10.1038/s41567-019-0650-1} {\bibfield  {journal} {\bibinfo  {journal} {Nat. Phys.}\ }\textbf {\bibinfo {volume} {15}},\ \bibinfo {pages} {1227} (\bibinfo {year} {2019})}\BibitemShut {NoStop}%
\bibitem [{\citenamefont {Keesling}\ \emph {et~al.}(2019)\citenamefont {Keesling}, \citenamefont {Omran}, \citenamefont {Levine}, \citenamefont {Bernien}, \citenamefont {Pichler}, \citenamefont {Choi}, \citenamefont {Samajdar}, \citenamefont {Schwartz}, \citenamefont {Silvi}, \citenamefont {Sachdev}, \citenamefont {Zoller}, \citenamefont {Endres}, \citenamefont {Greiner}, \citenamefont {Vuleti{\'c}},\ and\ \citenamefont {Lukin}}]{LukinKBMZ}%
    \BibitemOpen
    \bibfield  {author} {\bibinfo {author} {\bibfnamefont {A.}~\bibnamefont {Keesling}}, \bibinfo {author} {\bibfnamefont {A.}~\bibnamefont {Omran}}, \bibinfo {author} {\bibfnamefont {H.}~\bibnamefont {Levine}}, \bibinfo {author} {\bibfnamefont {H.}~\bibnamefont {Bernien}}, \bibinfo {author} {\bibfnamefont {H.}~\bibnamefont {Pichler}}, \bibinfo {author} {\bibfnamefont {S.}~\bibnamefont {Choi}}, \bibinfo {author} {\bibfnamefont {R.}~\bibnamefont {Samajdar}}, \bibinfo {author} {\bibfnamefont {S.}~\bibnamefont {Schwartz}}, \bibinfo {author} {\bibfnamefont {P.}~\bibnamefont {Silvi}}, \bibinfo {author} {\bibfnamefont {S.}~\bibnamefont {Sachdev}}, \bibinfo {author} {\bibfnamefont {P.}~\bibnamefont {Zoller}}, \bibinfo {author} {\bibfnamefont {M.}~\bibnamefont {Endres}}, \bibinfo {author} {\bibfnamefont {M.}~\bibnamefont {Greiner}}, \bibinfo {author} {\bibfnamefont {V.}~\bibnamefont {Vuleti{\'c}}},\ and\ \bibinfo {author} {\bibfnamefont {M.~D.}\ \bibnamefont {Lukin}},\ }\bibfield  {title} {\bibinfo {title} {Quantum kibble--zurek mechanism and critical dynamics on a programmable rydberg simulator},\ }\href {https://doi.org/10.1038/s41586-019-1070-1} {\bibfield  {journal} {\bibinfo  {journal} {Nature}\ }\textbf {\bibinfo {volume} {568}},\ \bibinfo {pages} {207} (\bibinfo {year} {2019})}\BibitemShut {NoStop}%
\bibitem [{\citenamefont {Yi}\ \emph {et~al.}(2020)\citenamefont {Yi}, \citenamefont {Liu}, \citenamefont {Jiao}, \citenamefont {Zhang}, \citenamefont {Zhang},\ and\ \citenamefont {Chen}}]{PhysRevLett.125.260603}%
    \BibitemOpen
    \bibfield  {author} {\bibinfo {author} {\bibfnamefont {C.-R.}\ \bibnamefont {Yi}}, \bibinfo {author} {\bibfnamefont {S.}~\bibnamefont {Liu}}, \bibinfo {author} {\bibfnamefont {R.-H.}\ \bibnamefont {Jiao}}, \bibinfo {author} {\bibfnamefont {J.-Y.}\ \bibnamefont {Zhang}}, \bibinfo {author} {\bibfnamefont {Y.-S.}\ \bibnamefont {Zhang}},\ and\ \bibinfo {author} {\bibfnamefont {S.}~\bibnamefont {Chen}},\ }\bibfield  {title} {\bibinfo {title} {Exploring inhomogeneous kibble-zurek mechanism in a spin-orbit coupled bose-einstein condensate},\ }\href {https://doi.org/10.1103/PhysRevLett.125.260603} {\bibfield  {journal} {\bibinfo  {journal} {Phys. Rev. Lett.}\ }\textbf {\bibinfo {volume} {125}},\ \bibinfo {pages} {260603} (\bibinfo {year} {2020})}\BibitemShut {NoStop}%
\bibitem [{\citenamefont {Ho}\ and\ \citenamefont {Zhou}(2007)}]{PhysRevLett.99.120404}%
    \BibitemOpen
    \bibfield  {author} {\bibinfo {author} {\bibfnamefont {T.-L.}\ \bibnamefont {Ho}}\ and\ \bibinfo {author} {\bibfnamefont {Q.}~\bibnamefont {Zhou}},\ }\bibfield  {title} {\bibinfo {title} {Intrinsic heating and cooling in adiabatic processes for bosons in optical lattices},\ }\href@noop {} {\bibfield  {journal} {\bibinfo  {journal} {Phys. Rev. Lett.}\ }\textbf {\bibinfo {volume} {99}},\ \bibinfo {pages} {120404} (\bibinfo {year} {2007})}\BibitemShut {NoStop}%
\bibitem [{\citenamefont {Capogrosso-Sansone}\ \emph {et~al.}(2007)\citenamefont {Capogrosso-Sansone}, \citenamefont {Prokof'ev},\ and\ \citenamefont {Svistunov}}]{PhysRevB.75.134302}%
    \BibitemOpen
    \bibfield  {author} {\bibinfo {author} {\bibfnamefont {B.}~\bibnamefont {Capogrosso-Sansone}}, \bibinfo {author} {\bibfnamefont {N.~V.}\ \bibnamefont {Prokof'ev}},\ and\ \bibinfo {author} {\bibfnamefont {B.~V.}\ \bibnamefont {Svistunov}},\ }\bibfield  {title} {\bibinfo {title} {Phase diagram and thermodynamics of the three-dimensional bose-hubbard model},\ }\href@noop {} {\bibfield  {journal} {\bibinfo  {journal} {Phys. Rev. B}\ }\textbf {\bibinfo {volume} {75}},\ \bibinfo {pages} {134302} (\bibinfo {year} {2007})}\BibitemShut {NoStop}%
\end{thebibliography}
\end{document}